\documentclass{aa}  
\usepackage{graphicx}
\usepackage[varg]{txfonts}
\usepackage{enumerate}
\usepackage{enumitem}
\usepackage{color}
\usepackage{natbib}
\newcommand{\xs}{{\sc X-shooter}\ } 
\newcommand{\xss}{{\sc X-shooter}'s\ }

\newcommand{\gt}[1]{>}

\begin{document}
\title{The Panchromatic High-Resolution Spectroscopic Survey of Local Group Star Clusters\thanks{Based on observations collected under program 084.D-1061 with \xs at the Very Large Telescope of the Paranal Observatory in Chile, operated by the European Southern Observatory (ESO).}}
\subtitle{I. General data reduction procedures for the VLT/X-shooter UVB and VIS arm}

\titlerunning{The Panchromatic High-Resolution Spectroscopic Survey of Local Group Star Clusters}
\authorrunning{Sch\"onebeck et al.}

\author{Frederik Sch\"onebeck\inst{1,\thanks{Member of the International Max-Planck Research School (IMPRS) for Astronomy \& Cosmic Physics at Heidelberg University and the Heidelberg Graduate School of Fundamental Physics (HGSFP).}}, 
Thomas H.~Puzia\inst{2}, 
Anna Pasquali\inst{1},
Eva K.~Grebel\inst{1}, 
Markus Kissler-Patig\inst{3}, 
Harald Kuntschner\inst{4},
Mariya Lyubenova\inst{5}, \and
Sibilla Perina\inst{2}
}

\institute{Astronomisches Rechen-Institut, Zentrum f\"ur Astronomie der Universit\"at Heidelberg, M\"onchhofstra\ss e 12-14, 69120 Heidelberg, Germany, \email{frederik@ari.uni-heidelberg.de}
\and
Institute of Astrophysics, Pontificia Universidad Cat\'olica de Chile, Av. Vicu\~na Mackenna 4860, Macul 7820436, Santiago, Chile
\and
Gemini Observatory, 670 N. A'ohoku Place, Hilo, HI 96720, USA
\and
European Southern Observatory, Karl-Schwarzschild-Str. 2, D-85748 Garching, Germany
\and
Kapteyn Astronomical Institute, 9700 AV Groningen, Netherlands}

\date{Received May 13, 2014 / Accepted September 4, 2014}

\abstract
     {}
     {Our dataset contains spectroscopic observations of $29$ globular clusters in the Magellanic Clouds and the Milky Way performed with VLT/\xs over eight full nights. To derive robust results instrument and pipeline systematics have to be well understood and properly modeled.~We aim at a consistent data reduction procedure with an accurate understanding of the measurement accuracy limitations. Here we present detailed data reduction procedures for the VLT/\xs UVB and VIS arm. These are not restricted to our particular dataset, but are generally applicable to different kinds of \xs data without major limitation on the astronomical object of interest.}
    {ESO's \xs pipeline (v1.5.0) performs well and reliably for the wavelength calibration and the associated rectification procedure, yet we find several weaknesses in the reduction cascade that are addressed with additional calibration steps, such as bad pixel interpolation, flat fielding, and slit illumination corrections.~Furthermore, the instrumental PSF is analytically modeled and used to reconstruct flux losses at slit transit.~This also forms the basis for an optimal extraction of point sources out of the two-dimensional pipeline product. Regular observations of spectrophotometric standard stars obtained from the \xs archive allow us to detect instrumental variability, which needs to be understood if a reliable absolute flux calibration is desired.}
    {A cascade of additional custom calibration steps is presented that allows for an absolute flux calibration uncertainty of $\lesssim10\%$ under virtually every observational setup, provided that the signal-to-noise ratio is sufficiently high. The optimal extraction increases the signal-to-noise ratio typically by a factor of $1.5$, while simultaneously correcting for resulting flux losses. The wavelength calibration is found to be accurate to an uncertainty level of $\Delta\lambda\!\simeq\!0.02$\,\AA.}
    {We find that most of the \xs systematics can be reliably modeled and corrected for. This offers the possibility of comparing observations on different nights and with different telescope pointings and instrumental setups, thereby facilitating a robust statistical analysis of large datasets.}


   \maketitle

\section{Introduction}
\xs is a single target, slit echelle spectrograph that covers a very wide spectral range ($3000\!-\!25000\,\AA$) at moderate resolution \citep{vernet2011}. It is the first of the European Southern Observatory's (ESO) second-generation instruments installed at the Very Large Telescope (VLT) on Cerro Paranal in the Chilean Atacama Desert.~By splitting the incoming beam into three independent arms (ultraviolet-blue, UVB: $3000\!-\!5900\,\AA$; visible, VIS: $5300\!-\!10200\,\AA$; near-infrared, NIR: $9800\!-\!25000\,\AA$), each one equipped with optimized optics, coatings, dispersive elements, and detectors, the design allows for extremely high sensitivity throughout the entire spectral range.~The individual spectrographs are attached to the instrument's central support structure, adding up to 2.5 metric tons mounted at the Cassegrain focus of VLT. Available slit widths range from 0.4\arcsec\ to 5.0\arcsec\ and can be specified independently for each arm.~The slit length, however, is fixed at $11\arcsec$.~Although designed as a traditional slit spectrograph, \xs also features an image slicer that reformats three adjacent lateral fields with a total field of view of $4\arcsec\!\times\!1.8\arcsec$ into a pseudo long slit of $12\arcsec\times0.6\arcsec$.~The echelle layout of the instrument produces $12\!-\!16$ curved and highly distorted spectral orders per arm, requiring a sophisticated reduction process for the best possible results. With a typical resolution $R\!\gtrsim\!10000$, \xs is armed with state-of-the-art technology that can address a vast number of astrophysical applications. The broad spectral coverage and its mount on the 8.2-meter VLT turn \xs into a powerful multi-purpose tool that offers the ability to constrain parameter spaces that otherwise would need to be studied with a number of different instruments. 

Recent achievements include detailed spectroscopic studies of young stellar objects \citep{youngstars1, youngstars2, youngstars3, youngstars4}, but also of the much older Galactic population of extremely metal-poor stars \citep{caffau2011, caffau2011nature, caffau2012}. \citet{spectrallib, spectrallib2, xsl2014} have started to compile an empirical stellar \xs spectral library (XSL) with more than 600 stars (including 100 asymptotic giant branch stars), taking advantage of \xss unique capability to observe all frequencies from the ultra-violet (UV) atmospheric cutoff at $\sim\!3200\,\AA$ to the near-infrared (NIR) out to about $\sim\!2.5\,\mu$m in one exposure, thereby circumventing cross calibrations between different instruments with different technical specifications. 

In the field of extragalactic astronomy, where the target redshift, $z$, is often unknown prior to spectroscopic observations, \xss capabilities likewise offer tremendous benefits. For example, The Optically Unbiased GRB Host ({\sc TOUGH}) survey \citep{tough1} uses the wide spectral range to reliably measure the redshifts of 19 gamma-ray-burst host galaxies \citep{tough2}, which turns this instrument into a perfect showcase for studying faint star-forming galaxies at high redshifts. \citet{blackhole} used \xs to derive black hole mass estimates and emission-line properties for $z\!>\!6.5$ quasars, including the highest redshift quasars known to date (at $z\!\simeq\!7.1$) and \citet{quiescent} investigated the growth of massive, quiescent galaxies from $z\!\approx\!2$ to the local universe by directly probing their dynamical masses.

Our dataset contains integrated-light \xs observations of 29 globular clusters in the Milky Way and the Magellanic Clouds (ESO program 084.D-1061) inspired by the work of \cite{puzia02}.~The detailed description of the observing mode, sample selection, and first analyses will be presented in the next paper in this series.~Here, we discuss the general data quality of \xs and the accompanying pipeline for the UVB and VIS arms.~Since our dataset contains over $400$ raw frames and $\sim\!1500$ calibration frames, we attached great importance to consistency and reproducibility of our analysis procedures.~To characterize the data quality we use a significant number of \xs observations from the ESO archive\footnote{\url{http://archive.eso.org/eso/eso_archive_main.html}}.~We mainly concentrate on regularly acquired standard star observations (primarly of the white dwarf GD71), as they form an excellent tool to search for hidden systematic effects and to monitor the temporal stability of the instrument. We carefully analyze the result of each pipeline step and -- if necessary -- adjust the available parameters to optimize the output. In case of unsatisfactory results or lack of treatment of instrument systematics by the ESO pipeline, we implemented the necessary modifications to the standard calibration steps with additional custom code. The methods presented in this work are not limited to our particular dataset of globular clusters. They are ready to be applied to a wide variety of astronomical objects without major modifications.

This paper is organized as follows: We start with an overview of the reduction cascade in Section \ref{sec:reduction_cascade}. Calibration steps executed prior to invoking the ESO pipeline are explained in Section \ref{sec:pre_pipe}, whereas Section \ref{sec:pipe} briefly summarizes the steps executed inside the ESO pipeline. Most of our calibrations, however, are applied to the spectra after the pipeline rectification procedure, and are explained in Section \ref{sec:post_pipe}.~We conclude with a summary of our procedures and findings in Section~\ref{sec:summary}.

\section{Overview of the data reduction cascade}
\label{sec:reduction_cascade}
The entire cascade of reduction steps is executed with a global script that first sorts all calibration frames for each observing night based on header information and subsequently runs the set of calibration recipes on the respective frames in a predefined sequence. The script is designed to operate fully automatically except for the flux calibration part, where the sensitivity function computation can be performed in interactive mode. Our code can likewise handle extended objects and point sources and accepts all three observation (OB) modes offered by {\sc X-shooter}: \emph{OFFSET} (object and background are imaged onto different frames), \emph{NOD} (object is observed multiple times at different slit positions), and \emph{STARE} (object and background are observed only once within the same frame).~Throughout this paper we have used version 1.5.0 of the ESO \xs pipeline \citep{modigliani2010}, although we obtain similar results with version 1.3.7.~We were not able to test versions 2.0 or later; these versions may include any of the described reduction and analysis routines used in this work. Therefore, the presented results are not necessarily valid for these newer pipeline releases (v2.4 as of June 2014).~This paper covers the reduction of UVB and VIS data, and we expect that most of the presented results will be applicable to the NIR arm without great modifications.~We plan to describe the implementation of these reduction tasks in an upcoming paper of this series.

Our calibration recipes are implemented in \emph{IDL (Interactive Data Language)} and \emph{PyRAF (IRAF\footnote{IRAF is distributed by the National Optical Astronomy Observatories, which are operated by the Association of Universities for Research in Astronomy, Inc., under cooperative agreement with the National Science Foundation.} command language based on Python)} and, together with the ESO pipeline, are part of a \emph{Python} script that runs each step in a predefined order. Generally, the data reduction sequence can be categorized in three steps:~1) calibrations performed on the raw frames (pre-processing), 2) steps executed within the code provided by ESO (pipeline), and 3) calibrations applied to the rectified spectra (post-processing).~An overview of the calibration sequence is given in Table~\ref{tbl:pipe_steps}, where the presented order corresponds to the order of execution.~We note that we do not include the telluric correction in this work, but defer its detailed description to subsequent papers when we describe the data reduction and analysis steps of the NIR arm.~A telluric correction for \xs data has been discussed in a recent paper by \cite{xsl2014}.

\begin{table*}[t]
\centering
 \begin{tabular}{ p{2cm} p{8cm} p{2.5cm} p{2.5cm} cccc}
 \hline
 \hline
  \textbf{Section} & \textbf{task description} & \textbf{data type} & \textbf{OB mode} & \\
  \hline\\
  \multicolumn{5}{c}{\textbf{pre-processing reductions}}\\
  \hline
  \ref{sec:bias_noise} & pick-up noise elimination in bias frames & all & all & \\
  \ref{sec:io_background_model} & inter-order background and pick-up noise modeling in science frames & all & all & \\
  \hline\\
  \multicolumn{5}{c}{\textbf{pipeline reductions}}\\
  \hline
  \ref{sec:bias} & bias subtraction & all & all &\\
  \ref{sec:crh} & cosmic-ray hits removal & all & all\\
  \ref{sec:wavelength} & wavelength calibration and extraction (rectification) & all & all \\
  \hline\\
  \multicolumn{5}{c}{\textbf{post-processing reductions (on rectified spectra)}}\\
  \hline
  \ref{sec:error} & error map rescaling & all & all &\\
  \ref{sec:io_sub} & illumination background and pick-up noise subtraction& all & all \\
  \ref{sec:bp} & bad pixel interpolation & all & all &\\
  \ref{sec:ff} & flat fielding & all & all\\
  \ref{sec:illu_cor} & illumination correction & all & all &\\
  \ref{sec:nodding} & nodding & point source & \emph{NOD} & \\
  \ref{sec:sky_sub} & sky subtraction & point source & \emph{STARE} & \\
  \ref{sec:opt_extr} & optimal extraction & point source & all &\\
  \ref{sec:fluxcal} & absolute flux calibration and order merging & all & all &\\
  \ref{sec:wavelength_update} & fine-tuning of the wavelength calibration & all & all &\\
  \hline
  \hline
  \end{tabular}
  \caption{Calibration sequence for \xs data. Column 1 shows the sections under which the various calibration steps (Col. 2) are described in this paper. The data types (\emph{point source, all}) for which the recipes are applicable are given in the Col. 3. Column 4 contains information about the OB modes (\emph{STARE, NOD, OFFSET, all}) that can be handled by the recipes.}
  \label{tbl:pipe_steps}
\end{table*}
  
Most of the calibrations were implemented after the rectification for reasons of technical simplicity and user friendliness. Nevertheless, we want to emphasize that for a fully consistent error propagation a rectification and resampling of the curved echelle orders should be avoided, as the accompanying kernel convolution interpolation does not preserve the fundamental noise characteristics of the data.~Despite this, we decided to maintain the overall \emph{modus operandi} predetermined by the ESO pipeline (i.e.,~extracting and rectifying the echelle orders) and provide reasonable error estimates for all additionally required steps.~Depending on the data type (extended object or point source) and OB mode (\emph{STARE, NOD, OFFSET}) only a subset of the presented recipes is executed. For most of the mentioned recipes, our code produces various control plots, with which the quality of the calibration performance can be evaluated and optimized, if required.

\section{Pre-processing reductions}
\label{sec:pre_pipe}
Several calibrations have to be performed directly on the raw frames, as the underlying systematic effects are not limited to individual echelle orders, but affect the global count distribution of the entire CCD. 

\subsection{Pick-up noise elimination in bias frames}
\label{sec:bias_noise}
The UVB CCD of \xs is susceptible to pick-up noise in calibration and science frames, which manifests itself as an additional periodic pattern that fluctuates around the bias level with an amplitude of up to $\sim\!2$ counts distributed over various frequencies, with a phase that changes from frame to frame. The extrema are orientated along the $NAXIS1$ axis and have a typical spacing of approximately $10$ pixels. This alignment offers the possibility of removing the pick-up noise pattern by a one-dimensional Fourier filtering technique. For this, the raw bias frames are first checked for outliers (cosmic ray hits, bad pixels), which are then replaced by a locally estimated median.~In order to appropriately check for outliers, we implemented a $\kappa-\sigma$ clipping and carefully adjusted the involved parameters.~The box that is used to compute the local median and standard deviation is a vertical one-dimensional stripe of 51 pixels in length and the clipping threshold is chosen to be $8\,\sigma$.~The chosen box size represents a compromise that reliably eliminates both extended cosmic ray hits (CRH) and CCD blemishes, and at the same time is sufficiently small in order not to wash out any global trends in the overall bias structure.

Each column treated in this way is then {\sc Fourier}-transformed separately and the absolute values of the {\sc Fourier} modes are computed.~A $\kappa$-$\sigma$ clipping process is applied to the power spectrum that flags any dominant frequencies  that are $6\,\sigma$ outliers. Their real and imaginary parts are then replaced by the corresponding mean values of the adjacent modes, thereby preserving the amplitude and phase of the overall noise pattern. Back-transforming the clipped Fourier spectrum finally yields free bias frames that are free of pick-up noise, which can be used in subsequent processing steps.~Pick-up noise has not been detected in the VIS data for any of our reduced frames. Therefore, this calibration step is only applied to the UVB data. 

\subsection{Intraorder background and pick-up noise modeling in science frames}
\label{sec:io_background_model}
\begin{figure}[t]
\includegraphics[width=9cm, bb=25 0 630 316]{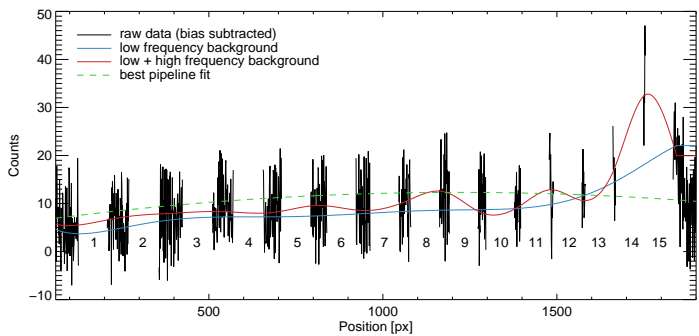}
\caption{Horizontal cut through a bias subtracted VIS raw frame of GD71. The blue curve shows the low order {\sc Chebyshev} fit and the red curve the superposition of the {\sc Chebyshev} fit and an additional spline fit. The best background fit obtained with the ESO pipeline is overplotted as a dashed green line. The gaps in the data (black curve) indicate the positions of the echelle orders. The respective order numbers are specified.}
\label{fig:io_back2}
\end{figure}

Unfortunately, the algorithm presented in Section \ref{sec:bias_noise} does not work for inhomogeneously illuminated CCDs with pronounced discontinuities in the global light distribution.~The cross-disperser of \xs produces strongly curved echelle orders that have separations of $5\!-\!70$ pixels, depending on the spectral order number. The inter-order (i.e.,~between the individual echelle orders) background count level, caused by a combination of the stray light, light diffusion inside the prism, and the change in order spacing, is a highly non-linear pattern that follows the illumination intensity of the adjacent echelle orders, which themselves are functions of both the spectral energy distribution (SED) and the blaze function.~The illumination background inside each order can only be reconstructed by propagating the illumination pattern between the orders with interpolation and/or fitting techniques.~In the inter-order background subtraction implemented in the \xs pipeline package (v1.5.0), the user has the possibility of fitting a global two-dimensional polynomial of low order to selected background regions.~This yields a smooth surface that follows the overall shape of the background counts distribution. However, it fails to reproduce small-scale effects like additional pick-up noise or enhanced stray light by line emission.~Furthermore, the VIS arm throughput shows a rapid increase between order 12 and its maximum at order 14, as well as a steep drop-off for even longer wavelengths, which cannot be modeled satisfactorily by the pipeline owing to the comparatively large correlation length of the fitted surface. An example plot demonstrating this issue is presented in Figure \ref{fig:io_back2}, where we show a horizontal cut through a GD71 VIS raw frame in which the echelle orders have been masked out so that the underlying illumination surface can emerge. The best fit obtained with the pipeline is overplotted in dashed green, which does not reproduce the scattered light component close to the strongly illuminated orders 14+15.~These shortcomings encouraged us to implement a subtraction technique that can reproduce both the global trend of the general background illumination as well as additional small-scale effects, such as pick-up noise and line emission. For this, we decided to model the background of each CCD row individually, thereby naturally including the horizontally aligned pick-up noise described in the previous section.

First, the two-dimensional raw frame is bias subtracted with a master bias frame, which is a median stack of a time series of bias exposures corrected for pick-up noise and is calculated within the original pipeline (see Section~\ref{sec:bias}).~The echelle order positions are then read out from the corresponding table that is produced during the wavelength calibration (see Section \ref{sec:wavelength}). Since diffraction at the slit ends is considerable in case of bright and extended objects, e.g.,~sky lines, we add an additional safety margin of up to 6 pixels on either side of the echelle order because additional diffraction counts are not to be included in the inter-order background level, as they only occur outside but not inside the echelle order. The actual size of the margin depends on the amplitude of the blaze function and the associated level of diffraction at the particular pixel position of interest, i.e.,~wider margins for more strongly illuminated orders.

Next, the statistical outliers in the inter-order sections are removed in a similar way to that described in Section~\ref{sec:bias_noise}, with $\kappa\!=\!4$ and a vertical box size of $151$ pixels, allowing for a better treatment of the significant number of cosmic ray hits.~We choose the vertical direction because the extent of the inter-order space decreases monotonically along the horizontal direction, approaching only a handful of pixels between some orders in the VIS arm.~Pixels masked by the echelle order apertures are excluded from the boxcar median estimate during the $\kappa-\sigma$ rejection, implying that the box size is reduced in the direct vicinity of the echelle orders if necessary.

In case of low signal-to-noise (S/N) data, in which the object only produces a very weakly illuminated background level (on average $<\!2$ counts per pixel), we found it sufficient to fit the inter-order sections of each row with a low-order polynomial. For this, the median of each inter-order section (per row) is computed, resulting in $n\!+\!1$ values for $n$ echelle orders.~Before the bluest order (1) and after the reddest order (UVB: 12, VIS: 15) we specify a rather narrow margin of 10 pixels for the median estimate, accounting for the fact that the information content on the illumination background decreases with increasing distance to the echelle orders. The section medians are subsequently fitted with a fifth-order {\sc Chebyshev} polynomial and the row-by-row fits are stored into a two-dimensional frame. The low polynomial order is chosen to ensure that even low S/N data with average background count levels of less than one can be robustly fitted and pick-up noise can be reliably detected even in the absence of strong stray light and light-diffusion components.

In cases where the inter-order background signal is on average $>\!2$ counts per pixel and follows the illumination distribution inside the orders, a two-step modeling approach is implemented.~We first mask out inter-order sections that typically show an enhanced level of background illumination (i.e.,~VIS orders 13,\,14,\,15), and fit the remaining sections row by row with an eighth-order {\sc Chebyshev} polynomial. The residuals between fit and data are checked for statistical outliers with a $2\,\sigma$-clipping and, if necessary, a second fitting iteration is performed without the sections not well modeled by the fit (e.g,.~due to line emission).~This yields a background model that is well adjusted to the smooth, low frequency component of the illumination surface.~Subsequently, the section medians of the residuals between data and low-order background model are interpolated with a third order spline function that accounts for localized line emission and large gradients in the throughput function.~This component (red minus blue fit in Figure~\ref{fig:io_back2}) is then added to the low-order estimate and the row-by-row models computed this way are stored into a global background model. Examples for the obtained low (blue) and low + high (red) frequency background components are shown in Figure~\ref{fig:io_back2} and present a considerably better fit than the one obtained with the built-in pipeline recipe (dashed green).~Similar to the treatment of the bias frames, the superposition is eventually {\sc Fourier} transformed column by column.~Any dominant modes are extracted and back-transformed into a separate frame, while the remaining spectrum (with the extracted modes replaced by the amplitudes of the adjacent channels; similar to Section \ref{sec:bias_noise}) is back transformed separately, and subsequently smoothed with a $7\!\times\!7$ pixel boxcar median to suppress possible fitting artifacts (i.e.,~random line-to-line fluctuations).

\begin{figure}[t]
  \begin{minipage}[b]{\linewidth}
  \centering
  \includegraphics[width=\textwidth]{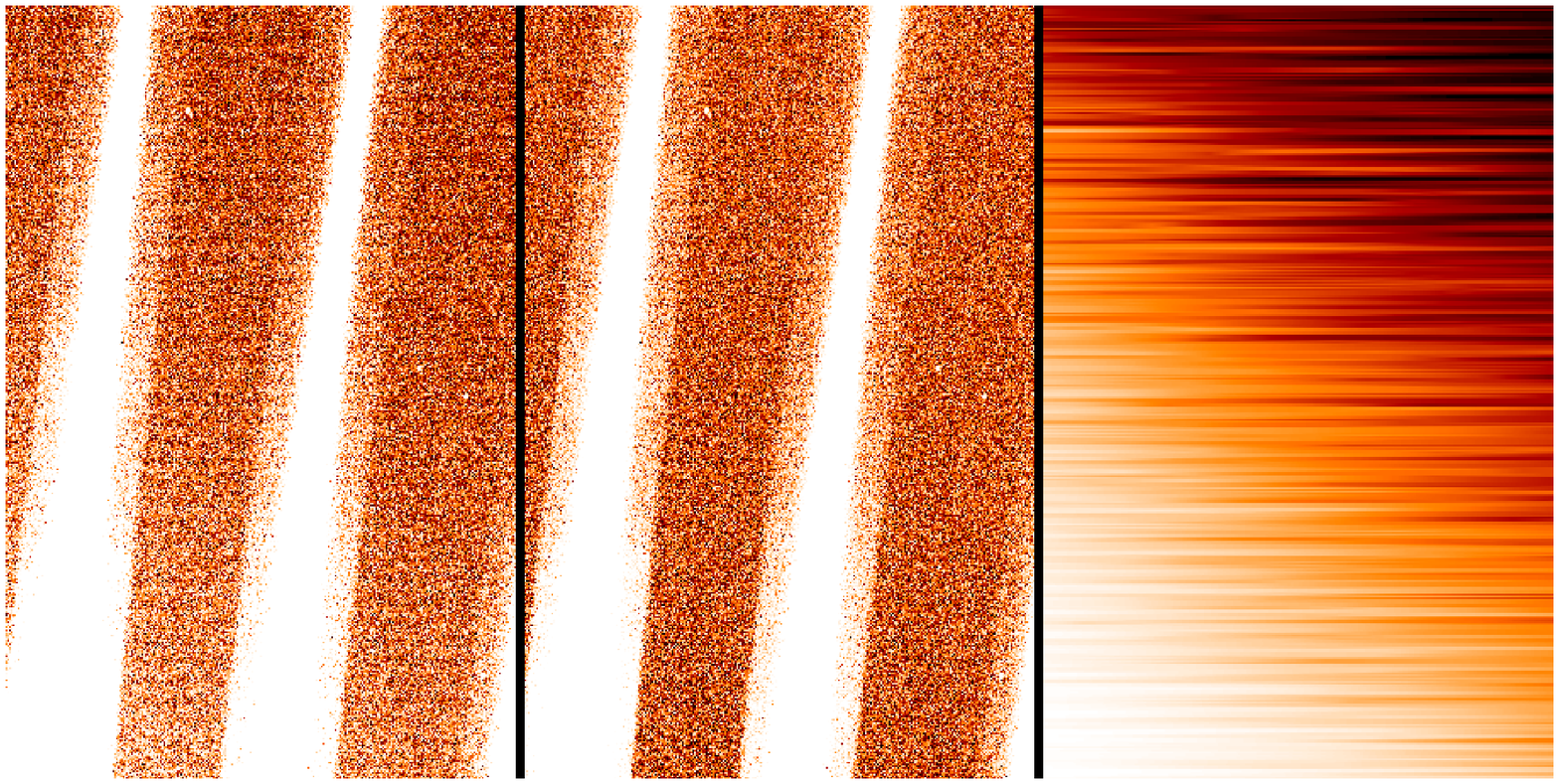}
 \end{minipage}
 \vfill
 \begin{minipage}[b]{\linewidth}
  \centering
  \includegraphics[width=\textwidth, bb= 70 98 613 143,clip]{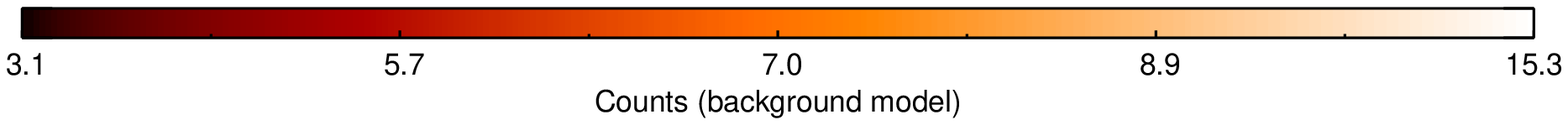}
 \end{minipage} 
 \caption{Inter-order background subtraction results for GD71. The left panel shows a section of a UVB raw frame (two illuminated echelle orders and the inter-order space between them), with illumination background (including pick-up noise). The central panel shows the same CCD section with applied inter-order background corrections. The corresponding illumination background model is presented in the right panel and its color scaling is illustrated in the bar at the bottom (bar valid only for right panel). The horizontally aligned pick-up noise shows up as a periodically alternating level of reduced and enhanced counts and is accurately fitted by the model. The color scales of all images have been histogram equalized.}
 \label{fig:io_back1}
\end{figure}

To demonstrate the two-dimensional performance of our approach with an emphasis on the correction for pick-up noise, the left and central panels of Figure \ref{fig:io_back1} show a before and after comparison of a GD71 UVB exposure using the inter-order background subtraction.~The right panel contains the corresponding two-dimensional background model with the characteristic pick-up noise pattern, which is significantly reduced after subtraction (central panel). 

The two-dimensional inter-order background model is stored in a \emph{FITS} file and the subtraction is performed after rectification (see Section \ref{sec:io_sub}), in order to correctly propagate the flux uncertainties into the rectified frame. We note that the ESO pipeline computes Poisson flux errors from the raw frame. 

\section{Pipeline data reduction steps}
\label{sec:pipe}
\subsection{Bias subtraction}
\label{sec:bias}
Typically, five bias frames corrected for pick-up noise (see Section~\ref{sec:bias_noise}) are specified as input for \emph{xsh\_mbias}, which generates a master bias frame and the associated error map. The master frame is a median stack of all supplied frames and the error values are set to the readout noise level.~The master bias frame serves as input for all subsequent pipeline recipes. A more detailed explanation is given by \citet{modigliani2010} and the corresponding pipeline manual\footnote{http://www.eso.org/sci/software/pipelines/}. 

\subsection{Removal of cosmic ray hits}
\label{sec:crh}
The cosmic ray hit rejection algorithm implemented in the \xs pipeline (v.1.5.0) is based on edge detection by Laplacian convolution \citep{vanDokkum} and is performed on the raw frames.~We obtained the best CRH rejection results with the following parameter settings: \emph{--removecrhsingle-sigmalim=2.0} (Poisson fluctuation threshold, default=$5.0$) and \emph{--removecrhsingle-flim=1.1} (minimum contrast between Laplacian image and fine structure image, default=$2.0$). 

To check for possible side effects we carefully compared the results of our parameter choice to the ones obtained with the default values. In particular, we focused on targets with bright emission lines in narrow slit setups, for which we expect the risk of confusion between CRH and line spread function to be the highest. We could detect no significant flux changes by the CRH rejection algorithm except for one case in which the $H\alpha$ transition was saturated, the resulting discontinuities were flagged, and the associated count values slightly altered. 

By contrast, the direct vicinity of bad pixels in highly exposed parts of the CCD form the only scenario for which we could systematically detect undesired flux changes for our choice of parameters (apart from true CRHs), which typically result in count drops $\gtrsim 20\%$ per affected pixel. This effect is particularly pronounced for the two bad pixel columns intersecting echelle order 5 of the VIS arm. Apparently, the prior information on bad pixel positions is not used during the CRH flagging, as otherwise the detection algorithm should have been adjusted accordingly to the underlying bad pixel map to avoid the above mentioned side effect. Unfortunately, as of v1.5.0 of the ESO \xs pipeline, the location of flagged CRH pixels is not directly propagated into the quality control maps. However, the CRH positions are stored in a separate \emph{FITS} extension, which we cross correlate with our master bad pixel map (see Section \ref{sec:bp} for a detailed description of its derivation and the subsequent use) to find any CRH-affected pixels in the direct vicinity of bad pixels. The threshold distance (with respect to the closest bad pixel) for a CRH position to be flagged as critical is set to $2$ pixels, whereas the standard two-dimensional Euclidean norm is used to compute the necessary separations. Only these CRH positions are inherited by our master bad pixel map and are later corrected for possible flux changes.~For all other locations we rely on the built-in CRH rejection routine to estimate and recover the original values. 

\subsection{Wavelength calibration and order extraction}
\label{sec:wavelength}
The wavelength calibration and order extraction process of both the UVB and VIS arm are based on a series of ThAr lamp and flat field lamp frames.~A first spectral format estimate is generated with two-dimensional Gaussian fits to the spectral lines of a single pinhole arc lamp exposure.~Crossmatching the obtained line positions with a reference line catalogue yields an optimization of the instrument's physical model, which is based on a ray-tracing algorithm for different wavelengths and slit positions \citep{bristowphysicalmodel}. Subsequently, the centers of the individual echelle orders are traced with a single pinhole flat lamp exposure. In order to obtain a full, two-dimensional wavelength solution that maps the pixel coordinates of the raw frame $(x,y)$ onto the physical grid of wavelength and slit position $(\lambda, s)$, an additional nine pinhole ThAr frame is acquired. Applying the same fitting techniques as for the single pinhole ThAr exposure, the emission peaks are accurately located and the physical model parameters refined.~With additional parameters from the science exposure FITS header, the model is finally adjusted to the ambient conditions of the observation and the raw data are rectified, extracted, and resampled to an equidistant wavelength grid with a kernel convolution interpolation.~To account for instrument flexure for off-zenith positions of the telescope and the accompanying shift in wavelength and slit position, additional flexure compensation frames are typically acquired and the appropriate calibration corrections performed automatically. A more detailed discussion on the achieved wavelength calibration accuracy is presented in Section \ref{sec:wavelength_update}.~Again, the reader is referred to \citet{modigliani2010} for a full description of all pipeline steps and to \citet{bristowphysicalmodel} for an extensive demonstration of the physical model capabilities.

\section{Post-pipeline reductions}
\label{sec:post_pipe}
We implemented most of the additional calibration steps on the pipeline-reduced and rectified echelle orders. This offers the possibility of working on equidistant, rectangular pixel grids that are calibrated in wavelength and slit position, which simplifies the technical complexity. However, the rectification process with its kernel interpolation introduces covariances that, if not properly modeled, potentially lead to inconsistent error estimates and thus to a vitiation of the spectral information content.~With this in mind, we paid particular attention to accurately quantifying uncertainties related to choosing post-pipeline corrections in order to achieve the highest possible data quality as well as a well characterized data structure with focus on reproducibility and manageability.

\subsection{Error map adjustments}
\label{sec:error}
A correct estimate of the uncertainties accompanying each observation and the subsequent reduction processes is crucial for any scientific analysis. We therefore carefully investigated each step of the error treatment and applied corrections to the pipeline implementation if necessary. 

\subsubsection{Error map rescaling}
We analyzed the accuracy of the error propagation in the pipeline rectification process and we found the implementation of the error calculation in the ESO pipeline to be incorrect (v1.5.0), although it is correctly described in the manual.~Based on the standard error estimate
\begin{equation}
 \sigma_{tot}[ADU] = \frac{\sigma_{tot}[e^{-}]}{g} = \frac{\sqrt{gF + (\sigma_r[e^{-}])^2}}{g} =  \sqrt{\frac{ F }{ g } + \left(\frac{\sigma_{r}[e^{-}]}{g}\right)^2}, 
 \label{eqn:error}
\end{equation}
where $F$ are the measured counts and $\sigma_r$ is the readout noise (specified in the file header), we noticed that the pipeline end product lacks the correct treatment of the gain $g$, which is given in units of $e^{-}/ADU$. For data that is dominated by Poisson noise, the rectified errors differ by one factor of the gain from the correct value, whereas in readout-noise-limited cases an additional gain dependent deviation occurs.~After extensive testing, we implemented a scaling that leads to the correct error form via
\begin{equation}
 \sigma_c = \frac{ \sigma_p \sqrt{\frac{F}{g} + \frac{\sigma_{r}^2}{g^2} + \frac{\sigma_{r,B}^2}{g^2}} }{g \sqrt{\frac{F}{g} + \sigma_{r}^2 + \sigma_{r,B}^2} },
\label{eqn:error_gain}
\end{equation}
where $\sigma_p$ is the incorrect error computed by the pipeline and $\sigma_c$ the error after correction.~The third term under both square roots of Equation (\ref{eqn:error_gain}) originates from the bias subtraction that is applied by the pipeline right before the rectification and takes into account the noise level $\sigma_{r,B}$ of the master bias frame.~However, because of the irreversibility of the kernel convolution during the rectification process, this remains only an approximate correction.~This incorrect error calculation implementation is an issue that affects all wavelengths and thus at least in cases for which $\sigma_r^2 \ll gF$, produces a constant deviation from the correct error estimates, which translates into a systematic S/N bias.

\subsubsection{Rebinning corrections}\label{sec:error_reb}
Because the ESO pipeline implements a simple {\it resampling} of the pixel information from the raw to the rectified frame (with a simplistic Gaussian error propagation of flux uncertainties and readout noise), the output grid does not encode the correct information about resampled pixel size information.~Here, we propose a different implementation that includes a {\it rebinning} of the raw grid, where the input fluxes are internally redistributed during the rectification and, hence, the associated errors rescaled according to the output pixel size, so that raw-frame S/N per wavelength unit is preserved in the rectified frame (modulo covariances of the interpolation convolution, see Section~\ref{sec:covarcon}).

We emphasize this resampling issue, because \xss dispersion relation varies by a factor of two in the covered wavelength ranges of the UVB and VIS arms.~In both arms, it typically starts at $0.1\,\AA\,$pix$^{-1}$ in the blue and ends at $0.2\,\AA\,$pix$^{-1}$ at redder wavelengths, with an almost linear increase in between. The average instrumental dispersion of each order is given in Col. 5 of Table~\ref{tbl:sensfunc}.

~During rectification, the pipeline interpolates the dispersion-variable input signal to an output grid of user-specified constant dispersion.~This interpolation is a basic resampling that does not account for the spectrally variable pixel size in the raw frame, but simply interpolates the detected counts to the desired output grid and applies a Gaussian propagation to the associated errors, irrespective of the size of the light collecting area at a given wavelength.~This leads to a conservation of pixel values (intensities) and a S/N scaling, which is independent of the chosen output dispersion. In fact, the measured counts on the raw images represent fluxes and already constitute an implicit integration (over the wavelength range and slit coordinate dimension covered by each pixel). Thus, since spectroscopy is typically interested in (differential) distribution functions, the input pixel sizes need to be traced and propagated accordingly to the error map. While these considerations theoretically also hold for the cross-dispersed dimension, any change in pixel size between raw and rectified spectrum essentially corresponds to a change in spectral dispersion, as \xss cross-dispersed pixel size ($\sim\!0.16\arcsec$/pixel) remains almost constant for all orders in the UVB and VIS arm.

To measure the input pixel size, we constructed an artificial raw frame with a constant count level and rectified it with the keyword \emph{rectify-conserve-flux=TRUE} (default \emph{FALSE})\footnote{Using the pipeline with either option (\emph{rectify-conserve-flux=TRUE} or \emph{FALSE}) does not treat the flux and variance values consistently.}.~When normalized by the input count level, the pixel values $\beta$ in the obtained rectified spectrum represent the relative output pixel size with respect to the true pixel size on the raw frame, i.e., 
\begin{equation}
 \beta = \frac{A_o}{A_i}, 
  \label{eqn:dispersion}
\end{equation}
where $A_{i,o}$ are raw (input) and rectified (output) pixel sizes. Dividing each pixel of the rectified error map by $\sqrt{\beta}$ finally yields an error spectrum with properly propagated information on the true light collecting area at each spectral position and ensures a conservation of the raw-frame S/N.

\subsubsection{Covariance considerations}\label{sec:covarcon}
In addition to the variable pixel size, covariance of the spectral data in the rectified images is another effect that has to be properly taken into account if an accurate estimate of the measurement uncertainties in later analysis steps is desired.~Pixel correlations are introduced by the kernel convolution process as part of the rectification procedure and the particular covariance contributions vary from pixel to pixel. To quantify the covariance we rectified a homogeneous noise frame (using the default kernel shape and size) and compared the standard deviations of the fluxes in the input and rectified frame. The standard deviation after rectification is lower by a factor of $1.22$, which translates into an average correlation length of $~1.5$ input pixels or $~1.5\times\beta$ output pixels (for an explation of $\beta$ see Section \ref{sec:error_reb}), respectively. Thus, the noise level inside the flux frame is underestimated and any further S/N calculations require the consideration of {\it both} the flux and variance frame.~We also stress that any subsequent analysis step needs to correct for the effective number of independent pixels $N_{\rm eff}$ when computing a reduced $\chi^2$ of any feature involving $N$ output pixels, which can be approximated by $N_{\rm eff}\approx N/(1.5\times\beta)$.~Although originally intended to be accessible for the user \citep{errorprop}, the per pixel covariance matrices are not output by the ESO pipeline and thus this piece of information is irreversibly lost and cannot be recovered from the data easily.\\

\subsection{Illumination background and pick-up noise subtraction}\label{sec:io_sub}
Once the two-dimensional wavelength and slit mappings are known, the previously constructed global background frame (see Section \ref{sec:io_background_model}) can be rectified in the same way as the associated science frame.~This is accomplished by copying the science header to the background model, so that both frames are pipeline-reduced with the same physical model parameters. This ensures a correct error treatment, since the error maps for both the science and the background frame are computed during the rectification (consisting of read-out noise and Poisson error), rescaled based on the corrections of Section~\ref{sec:error}, and can thus be propagated accordingly during the subtraction. While the fluxes are simply subtracted from each other, the errors of both frames are summed in quadrature, i.e.,
\begin{equation}
 \sigma_{tot} =\sqrt{\sigma_s^2 + \sigma_b^2},
 \label{eqn:gaussian_prop}
\end{equation}
where $\sigma_s$ and $\sigma_b$ are the rectified errors of the science frame and background frame, respectively. \\

\subsection{Bad pixel interpolation}
\label{sec:bp}
Since the bad pixel flagging and correction (based on the master bad pixel map computed during the master bias creation) did not work reliably with the \xs pipeline v1.5.0 for our dataset, we developed a routine that reliably flags bad pixels first, and then fits over the affected science regions to reconstruct the object signal. For reasons of technical simplicity (orthogonality of the $\lambda-slit$-coordinate system), we correct for bad pixels on the rectified spectra, which requires that the bad pixel information on the raw frame be accurately traced through the rectification process. 

In a first step a master bad pixel map is created. It is based on two median stacks of {\it linearity} flat fields (homogeneously illuminated frames), each set with its own exposure time. Both stacks are normalized by their respective exposure times and then divided by each other, resulting in a frame in which pixels with a strong non-linear response show significant deviations from unity.~For each pixel in this ratio frame the local median and standard deviation in a box with a side length of 25 pixels are computed. A $4\,\sigma$ detection threshold is applied to flag outliers, which are considered to be bad pixels and whose values are set to one in the master bad pixel map. Accordingly, good pixels are set to zero.

By appropriately updating all necessary header information, the master bad pixel map is rectified and extracted in exactly the same way as the science frames, ensuring that the information on the bad pixel positions is accurately traced and matches the two-dimensional mapping $(x,y \rightarrow slit, \lambda)$ of the science exposure. The kernel convolution interpolation that is applied during the rectification introduces pixel correlations (see Section \ref{sec:covarcon}) that map the original binary distribution of pixel values $\{0,1\}$ in the master bad pixel map onto the full range of $[0,1]$. We define the badness $b$ of a pixel as how much the rectified bad pixel map deviates from zero and refer to this badness map whenever we want to check the quality of the pixels in the rectified science output.

\begin{figure}[t]
\includegraphics[width=\linewidth]{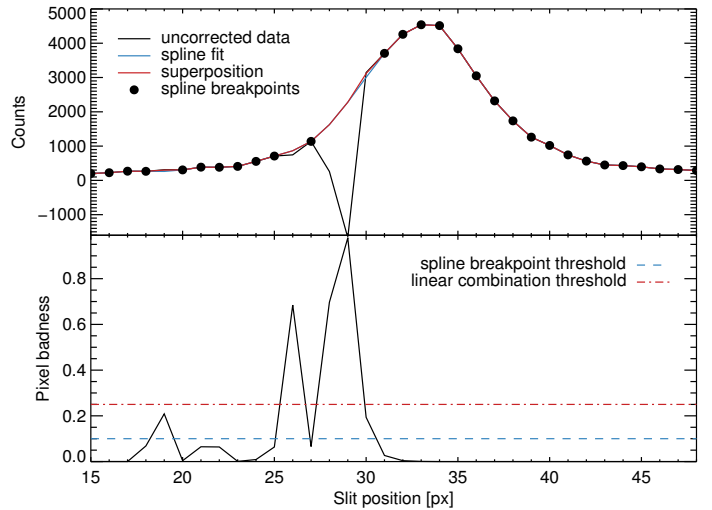}
\caption{Cut through rectified echelle order five of GD71 (VIS), which is intersected by two adjacent bad columns on the raw frame. The top panel shows the uncorrected PSF (black curve), the weighted spline fit (blue curve) and the weighted linear combination of both (red curve). The solid circles show the grid locations that are used as break points for the spline fit. The weighting factors are based on the pixel badness as depicted in the bottom panel. The thresholds for the spline break points (blue dashed line) and the weighted linear combination (red dash-dotted line) are overplotted. }
\label{fig:bp_fit}
\end{figure}

In order to correct the science observation for artifacts introduced by non-linear pixels, each wavelength bin of the science flux frame is fitted separately with cubic b-splines\footnote{IDL source code: \url{http://www.sdss3.org/svn/repo/idlutils/tags/v5_5_5/pro/bspline/bspline_fit.pro}} along the cross-dispersed direction on the rectified output (see Figure \ref{fig:bp_fit}). The break points for the spline fit are defined only at pixels where $b\leq0.1$ and pixels with $b>0.1$ are excluded from the fit.
Finally, a weighted linear combination $r$ of the uncorrected data $d$ and the corresponding fit $f$ is computed, 
\begin{equation}
 r = (1-b)\,d + b f, 
\end{equation}
however, only for pixel values for which  $b\leq0.25$. If $b > 0.25$ the comprised count value is considered to be too uncertain and the corresponding value of the spline fit is taken instead, because we encountered cases in which some pixels systematically showed count values $< -10^3$ after rectification (and bias subtraction), even though the corresponding raw signal was similar to the bias offset level. Such artifical outliers can dominate the linear combination even for $b\approx1$ and, thus, we decided to introduce a threshold in $b$ above which the corresponding flux values are ignored.~We note that both the spline break point threshold and the linear combination threshold were adjusted on an empirical basis and work reliably for different kinds of illuminations and profile shapes. 

\begin{figure}[t]
 \begin{minipage}[b]{\linewidth}
  \centering
  \includegraphics[width=8cm]{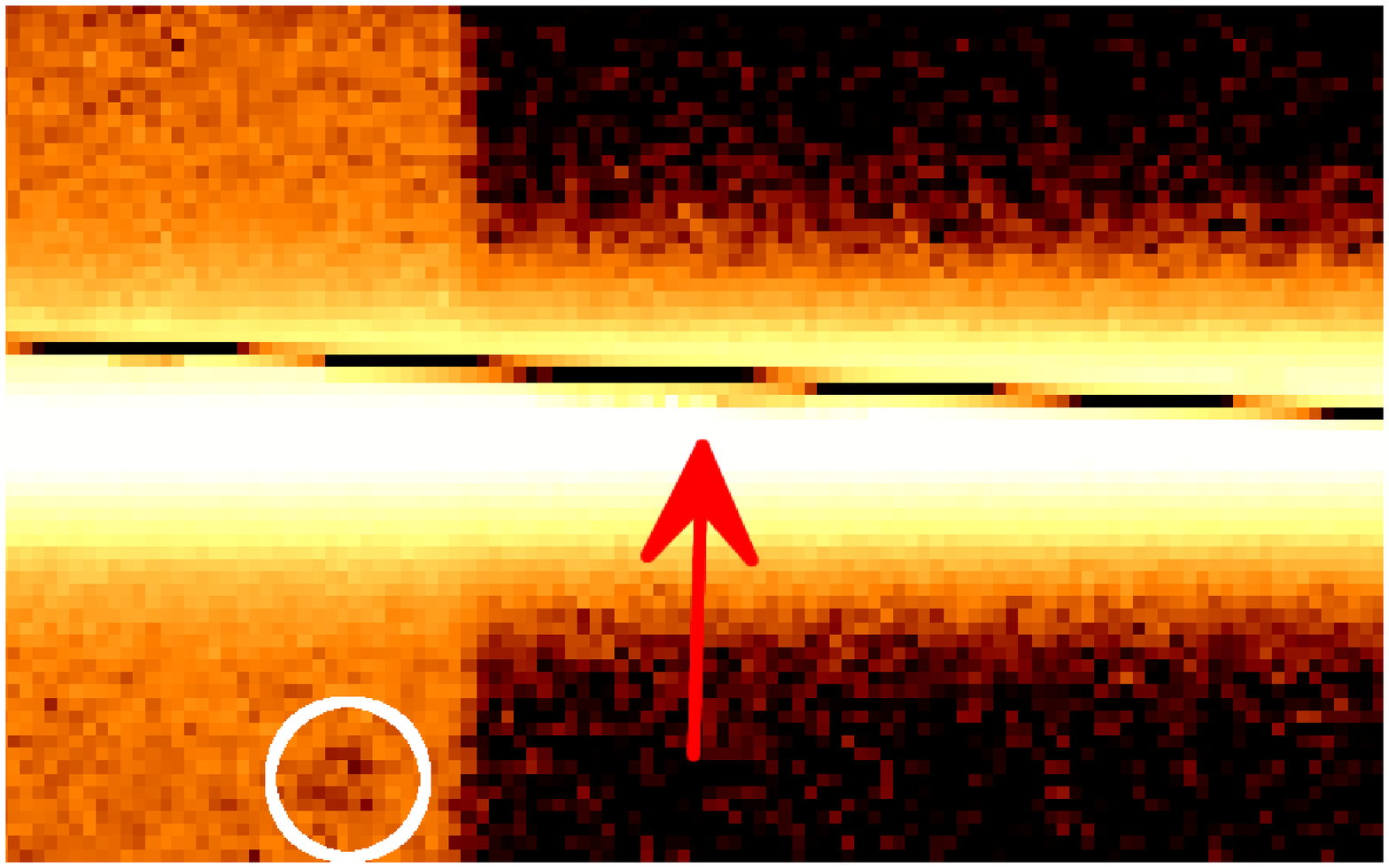}
  \end{minipage}
  \begin{minipage}[b]{\linewidth}
  \centering
  \includegraphics[width=8cm]{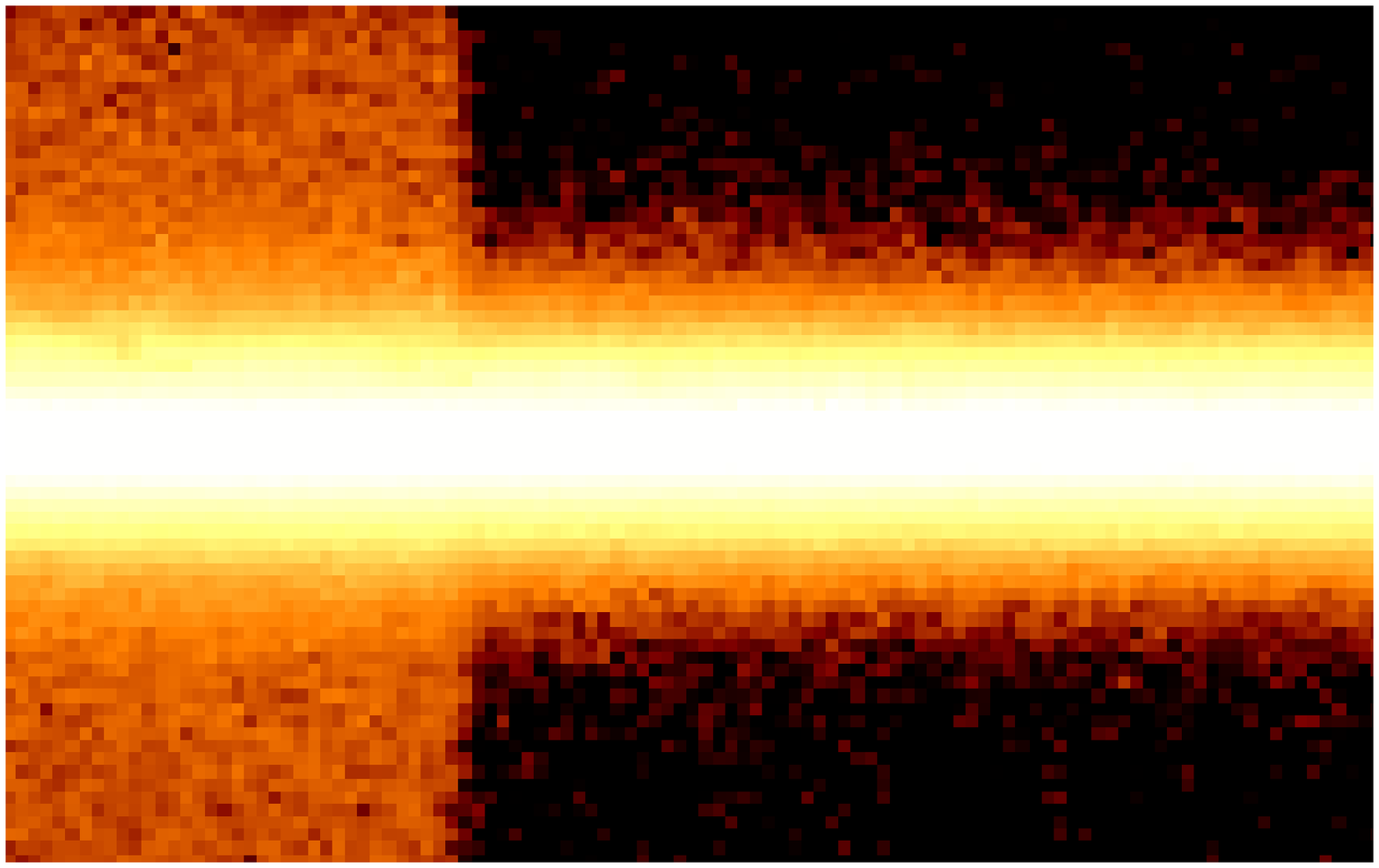}
  \end{minipage}
\caption{Zoomed-in section of the rectified echelle order five of GD71 (VIS) before (top) and after (bottom) application of the bad pixel interpolation. The artifacts of the bad columns 853 and 854 in the raw frame are visible as a dark stripe intersecting the illuminated echelle order. The area of reduced counts below the diagonal bad pixel stripe (indicated by the red arrow) is a consequence of the CRH rejection explained in Section~\ref{sec:crh}.~The cold spot at the lower left of the shown CCD section (marked by the white circle) is also reasonably corrected by our algorithm. The spectral dispersion direction runs horizontally along the x-axis. The plotted wavelength range is roughly $10\,\AA$ and the size along the cross-dispersed direction is $\sim\!4\arcsec$.~A vertical slice through these spectra is shown in Figure \ref{fig:bp_fit}.}
\label{fig:bp_data}  
\end{figure}

Choosing splines for the reconstruction of bad pixel affected signals has the advantage that no further information on the spatial PSF of the data is required. However, there is a risk that the original light distribution may not be optimally recovered.~We found our method to be robust for reasonably concentrated bad pixel regions ($\la\!3$ pix), which is the case for all our data.~To illustrate the general performance of our correction method, we show a before and after comparison in Figure \ref{fig:bp_data} for a section of echelle order five of GD71 (VIS), which is intersected by two bad columns on the raw frame ($x\!=\!853,854$).~The imprints of the bad pixels are significantly reduced; however, we still recommend inspecting problematic detector regions manually with the help of the rectified master bad pixel map and updating the correction function as necessary.

\subsection{Flat fielding}
\label{sec:ff}
Flat fielding is the next step that is applied to the bad pixel corrected data.~As we want to correct only for pixel-to-pixel quantum efficiency variations, the general shape of the flat field spectral energy distribution, which generally needs to be considered temporally variable (changes in temperature/voltage of the flat lamp), should be removed first, as otherwise it would be imprinted into the sensitivity function of the instrument. This raises the implicit need to reduce all frames with the same flat field image. By contrast, the \xs pipeline (v1.5.0) implements the flat fielding on the raw frames without any further treatment of the flat lamp exposure, except for a median stack and general wavelength-independent renormalization of the counts. Using such a treatment also means that any line signals emitted from the flat lamp are implicitly contained in the final flux-calibrated science spectrum.

\subsubsection{Quantification of flat field systematics}
Since we want to have the possibility of using one sensitivity function for our entire dataset on the one hand, but always use the flat field images closest in time to our respective science frames on the other hand, we implemented a flat fielding approach that is based on the rectified version of each image. This treatment allows for a simple and robust elimination of the arc lamp's spectral response by fitting a smooth polynomial along the spectral direction.~The drawback of this method, however, is related to the order in which additive and multiplicative steps are executed: ideally, the kernel convolution interpolation should be applied \emph{after} each raw pixel is flat fielded with its individual response. If rectified, such a raw-frame-based flat fielding implementation yields a flat fielded signal $f_{raw}$ and its associated noise model $\sigma_{raw}$ that are given by
\begin{equation}
 f_{raw} = \frac{\sum_i \frac{c_i}{r_i}{w_i}}{\sum_i w_i}, \, \,   \sigma_{raw} = \sqrt{\frac{\sum_i \left(\frac{\sqrt{c_i}}{r_i} \right)^2 w_i^2}{ \sum_i w_i^2 }},
 \label{eqn:ff_flux_pipeline}
\end{equation}
where $c_i$ are the raw pixel counts on which the convolution is performed, $w_i$ their respective kernel weights, and $r_i$ the individual raw pixel responses as determined from the flat field measurements.~In this simplified error estimation the only considered error component is Poisson noise and the flat fielding is assumed to be free of errors.

\begin{figure}[t]
\includegraphics[width=\linewidth]{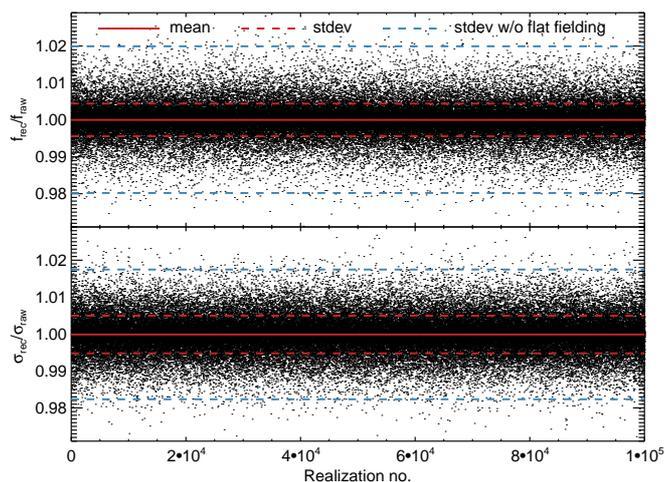}
\caption{Ratios of $10^5$ realizations of Equations (\ref{eqn:ff_flux_pipeline}) and (\ref{eqn:ff_flux_us}). The interpolation is performed with a Lanczos kernel for random locations inside the inner $\pm\,3\,\sigma$ of a Gaussian emission line with a FWHM of $3.5$ pixels (assuming Poisson noise). The pixel responses are modeled with a normal distribution with $\mu = 1$ and $\sigma = 0.02$. In both panels (top: flux ratios, bottom: error ratios) the solid red line shows the ensemble mean, the dashed red line its standard deviation, and the blue dashed line the standard deviation when the data are rectified but flat fielding is not applied afterwards.}
\label{fig:ff_accuracy}
\end{figure}

In our approach, however, the kernel interpolation is applied before the flat fielding, implying that the rectified pixel grid is flat fielded only with the kernel weighted average response, i.e.,
\begin{equation}
 f_{rec} = \frac{\sum_i c_i w_i}{\sum_i w_i} \left( \frac{\sum_i r_i w_i}{\sum_i w_i}\right)^{-1}, \, \, \sigma_{rec} = \sqrt{\frac{ \sum_i c_i w_i^2 }{ \sum_i w_i^2 }} \left(\frac{ \sum_i w_i r_i }{\sum_i w_i}\right)^{-1},
 \label{eqn:ff_flux_us}
\end{equation}
which produces slightly different results than Equation (\ref{eqn:ff_flux_pipeline}). To quantify the introduced numerical inaccuracy, we implemented both formulas in a one-dimensional Monte Carlo simulation with typical values for counts, kernel weights, and respones and compared their results. Figure~\ref{fig:ff_accuracy} shows both $f_{rec} / f_{raw}$ and $\sigma_{rec} / \sigma_{raw}$ for $10^5$ realizations. 

As input parameters for our simulation we selected values reflecting a typical \xs observation: a one-dimensional Gaussian emission line with a FWHM of $3.5$ pixels and a central count level of $10^4$ with an underlying continuum of $10^3$ counts. The exact count values for the grid were determined with a random number generator and a Poisson noise model was assumed. In addition, the individual pixel responses were drawn from a normal distribution with  $\mu = 1$ and $\sigma = 0.02$, the latter ones resembling a conservative upper limit on the expected pixel-to-pixel quantum efficiency variations. The rectification process was modeled with a standard Lanczos kernel with five lobes \citep{lanczos} and the count distribution was resampled with the kernel convolution at random grid locations inside the inner $\pm\,3\,\sigma$ of the emission line.

Flat fielding the convolved output grid (see Equation~\ref{eqn:ff_flux_us}) results in a $1\,\sigma$ uncertainty of $0.4\%$ with respect to the accurate scenario of flat fielding the input grid before resampling (Equation~\ref{eqn:ff_flux_pipeline}). This is lower than the Poisson noise of a single pixel of a typically illuminated flat field frame ($\sim\!1\%$). By contrast, omitting the flat fielding yields a $1\,\sigma$ deviation of $2.0\%$, which corresponds to the input response dispersion. The relative accuracy of the errors behaves in a similar way, with a $1\,\sigma$ deviation of $0.5\%$ when flat fielding the resampled output grid and $1.7\%$ if no flat fielding is applied at all. 

Choosing a flat count distribution of $10^4$ counts instead of an emission line profile for the input grid shows a deviation of only $0.03\%$ for the fluxes and $0.5\%$ for the errors. In both scenarios, this accuracy level is sufficient for all our purposes and thereby justifies implementing the flat fielding procedure on the rectified spectra.

\subsubsection{Removal of flat lamp emission lines}
In addition, working on rectified spectra also offers the possibility of easily removing emission lines of the flat field quartz lamp, which typically occur in the high energy range of the UVB arm. The sodium D2-lamp that is used to flat field the four bluest orders of the UVB arm of \xs shows a multiplicity of emissions lines with a typical strength of one to five percent with respect to the underlying continuum. Not removing them from the quartz lamp SED, will imprint their inverse profiles in the science frame as additional, artificial absorption features.

\begin{figure}[t]
  \includegraphics[width=\linewidth, bb=260 160 740 700, clip=]{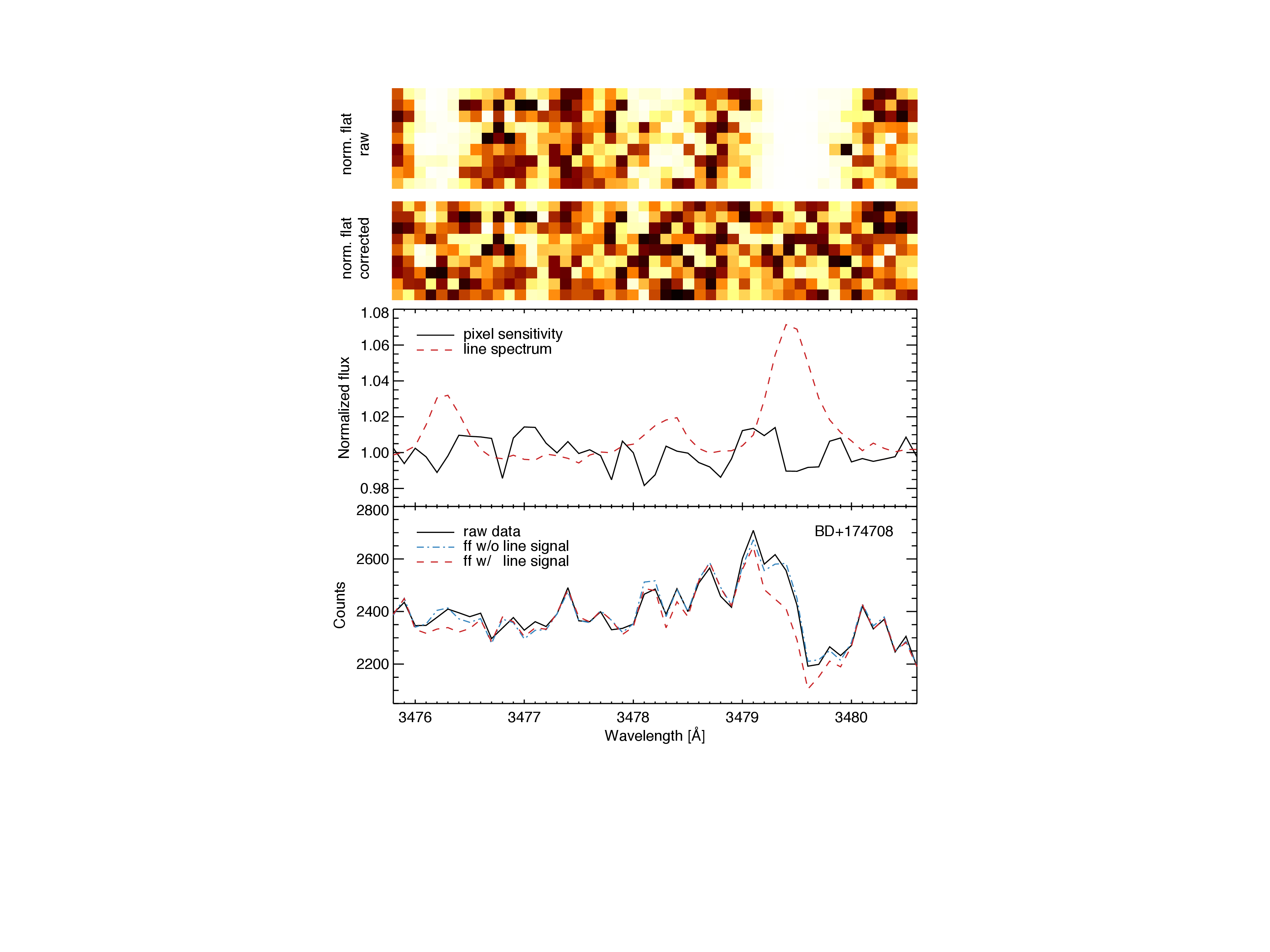} 
\caption{UVB flat field performance analysis. The top panel shows a zoomed-in section of the continuum-normalized SED of the D2 flat field lamp (echelle order four). The line emission is removed with an additional median along the cross-dispersed direction (second panel). In the third panel, we show the pixel-to-pixel quantum efficiency variations (solid black curve) at the central slit position and the corresponding line emission spectrum (dashed red curve).  
The bottom panel shows the SED of BD+174708 without flat fielding (solid black curve), flat fielded with the continuum-normalized D2 lamp SED (dot-dashed red curve), and flat fielded with its emission-line-free counterpart (dashed blue curve). The wavelength scales of the four panels are aligned so that the positions of the emission lines are the same for images and plots.}
 \label{fig:ff_comp}
\end{figure}

In our flat fielding implementation we first create a median stack of five flat field images, which is then wavelength calibrated and rectified in the same way as the science exposure. Subsequently, the bad pixels of the master flat are corrected as described in Section \ref{sec:bp}. The quartz lamp's spectral continuum and the response of the instrument are removed by row-wise fitting one-dimensional twelfth-order {\sc Chebyshev} polynomials along the spectral dispersion direction.~The impact of possible emission lines is removed by applying a $5\,\sigma$ clipping, together with a second fitting iteration. Normalizing by the continuum fit yields an intermediate solution that already carries direct information on the pixel-to-pixel variations, but still includes line signals, which can be removed by dividing each wavelength bin by its median value (see the two upper panels in Figure~\ref{fig:ff_comp}). The obtained quantum efficiency variations and the additional line emission signal are plotted in the third panel from the top in Figure~\ref{fig:ff_comp}. The necessity of correcting for quartz-lamp emission lines becomes obvious in the bottom panel in Figure~\ref{fig:ff_comp}, where we show the impact of such emission lines on the flat fielded SED of the spectrophotometric standard star BD+174708. If the correction for quartz-lamp emission lines is omitted, the stellar spectrum systematically deviates on a five percent level from the data that were correctly flat fielded.~This might lead to unexpected effects that are difficult to control if spectral features are located close to the emission features of the D2-lamp.

\subsection{Illumination correction}
\label{sec:illu_cor}
To achieve a relative flux calibration uncertainty of less than five percent for extended objects, we need to account for systematic illumination inhomogeneities along the slit. After the extraction and rectification process, the resulting slit illumination function is a combination of the instrument's imaging characteristics and the pipeline's ability to accurately trace and set extraction apertures around the curved echelle orders.

To measure the illumination function of \xs we use dedicated sky flat fields that were taken in addition to the regular calibration runs of the instrument (C. Martayan, private communication). The optical light path of a sky image should be similar to the ones of scientific observations except for instrumental flexure, which is unique to every observation and depends on telescope pointing and time (see Section \ref{sec:wavelength_update}).~By contrast, when the flat field quartz lamps are built into the instrument, we expect a different illumination pattern, rendering the quartz flats less useful to model the cross-dispersed illumination function of \xs in science operation mode.~The requirements imposed on the sky observations are a high S/N level, accurately compensated flexure, and no significant input signal gradient along the slit length of 11\arcsec.~Sky flat fields are only taken on special request, with the implication that we have only one set of suitable observations for our entire dataset.

\begin{figure*}[t]
\centering
 \begin{minipage}[b]{0.495\linewidth}
\includegraphics[width=9cm, bb=28 37 441 320,clip]{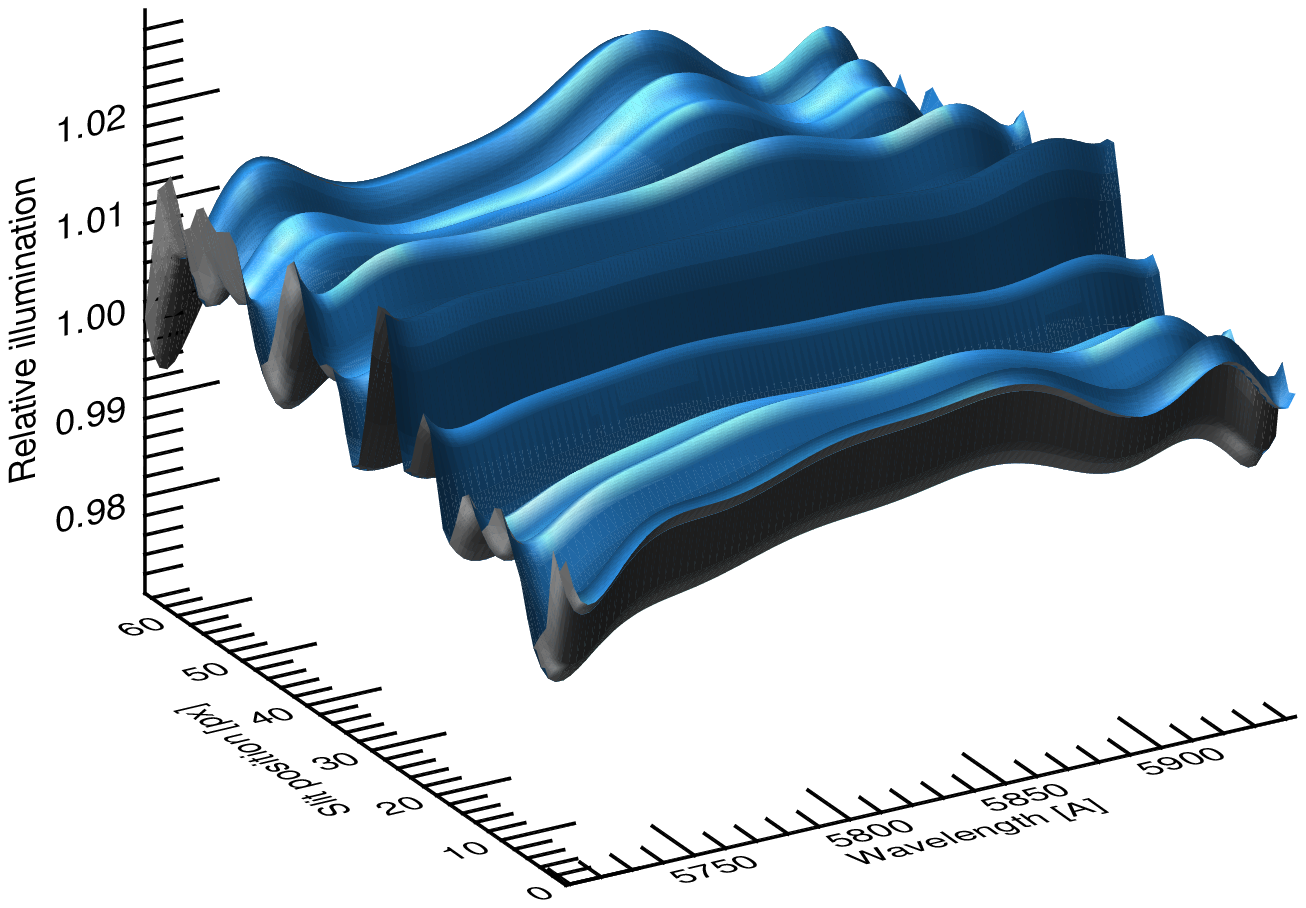}
  \end{minipage}
   \begin{minipage}[b]{0.495\linewidth}
\includegraphics[width=9cm, bb=28 37 441 320,clip]{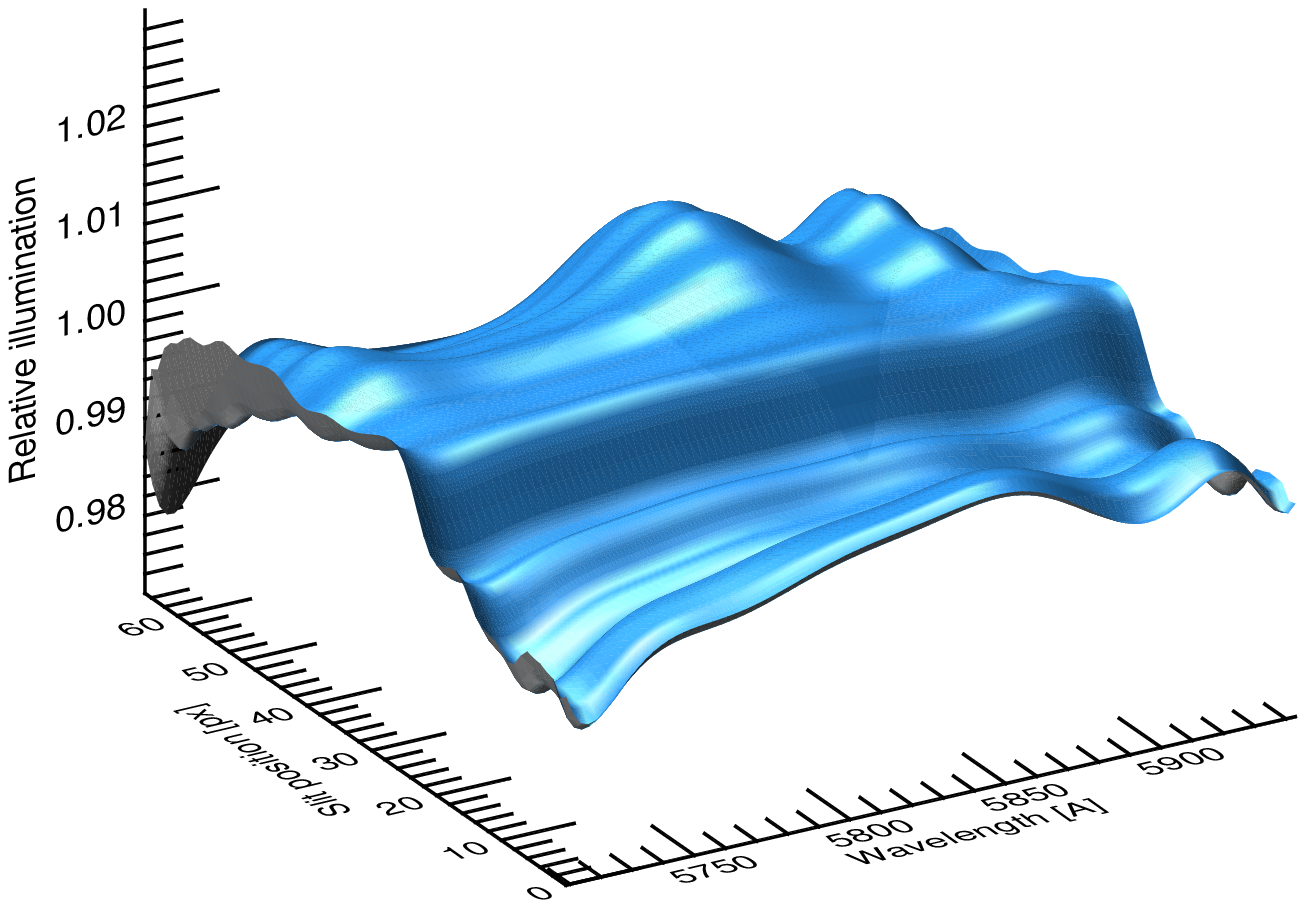}
  \end{minipage}
\caption{Two-dimensional illumination surfaces of echelle order three of the VIS arm for a $0.4\arcsec$ slit (left) and a $5.0\arcsec$ slit (right). The x-axis corresponds to the spectral dispersion direction, the y-axis to the cross-dispersed direction. The relative illumination is plotted on the z-axis. The maximum impact of the inhomogeneities is on the order of $\pm2.0\%$. The blue highlights and shadows were introduced to facilitate the visualization of the three-dimensional structure.}
\label{fig:illu_surface}
\end{figure*}

For two different slit widths in each instrument arm (UVB: $0.5\arcsec,5.0\arcsec$; VIS: $0.4 \arcsec,5.0\arcsec$), we rectified a sequence of five sky observations with their own flexure-compensated wavelength solution.~Subsequently, the output was corrected for bad pixels and flat fielded (see Sections \ref{sec:bp} and \ref{sec:ff}). We then normalized all cross-dispersed bins with respect to a weighted average of the central four rows, eliminating line emission and instrumental response.~The resulting spectra consist of illumination inhomogeneities and observational noise in the $(slit, \lambda)$ space.~The noise level was reduced by taking a weighted average using the corresponding error frames over the set of five exposures.~The illumination function can then be fitted with a high-order two-dimensional polynomial, yielding a smooth surface that follows any systematic gradients in the illumination pattern. Since each slit size has its own impact on the illumination function through different slit edge inhomogeneities and diffraction patterns, the above steps have to be executed separately for each slit size used. The left panel of Figure~\ref{fig:illu_surface} shows one example of the illumination surface for echelle order three of the VIS arm for a $0.4\arcsec$ wide slit.~The maximum amplitude of the illumination variations is on the order of $\pm2.0\%$ on a cross-dispersed length scale of $\sim\!5$ pixels. Furthermore, there are additional long-scale variations of comparable amplitude that produce smooth illumination gradients along both the entire slit length and the full spectral range. 

For the narrow slit setup, the above mentioned numbers are typical amplitudes for any echelle order on both tested instrument arms. For the $5.0\arcsec$ wide slit, we find that the small-scale fluctuations are considerably decreased and only the larger scale gradients among both cross-dispersed and spectral direction remain (see right panel of Figure~\ref{fig:illu_surface}).~Thus, small-scale variations may be caused by slit edge inhomogeneities, as those will have a much greater fractional impact on smaller slit widths, and/or moving dust aberrations. To distinguish between static or temporally varying systematics, additional calibration frames would be required.

In general, the observed illumination inhomogeneities are of rather small amplitude, yet, in order to achieve a relative systematic flux calibration uncertainty below a few percent, a proper illumination correction is recommended. 

\subsection{Nodding}
\label{sec:nodding}
Measurements conducted in \emph{NOD} mode typically consist of two (or multiples of two) observations of the same target, between which the telescope pointing is slightly offset so that the object is imaged onto different slit positions for each observation. By co-adding the frames in the correct way, the sky signal is naturally removed, circumventing the need for a tedious sky modeling process. Yet, for the nodding technique to work accurately the sky signal needs to stay constant between the individual exposures, rendering this mode practically useless for any observations performed in twilight (e.g., standard star observations, see Section~\ref{sec:sky_sub}) when the sky signal might vary up to $20\%$ within ten minutes. 

The \xs pipeline features a built-in treatment of nodded data, in which the co-addition is performed for the raw frames, yielding a rectified combined output. Nevertheless, the calibration steps presented in this paper are not applicable to these kinds of data in the strict sense, as multiplicative corrections are not commutable with summations, i.e.,
\begin{equation}
  \sum_i \frac{F_i}{S_i} \neq \frac{\sum_i F_i}{\sum_i S_i},                                                                                                                                                                                                                                                                                                                                       \end{equation}
where $F_i$ are the detected counts and $S_i$ arbitrary correction factors of the pixels of interest.~A similar reasoning was already presented in Section~\ref{sec:ff}, where we describe how the kernel convolution of the rectification affects the accuracy of the flat fielding process.

We resolve this by rectifying and calibrating each sequential frame independently. All post-pipeline steps described in Sections~\ref{sec:error} to \ref{sec:illu_cor} are applied to each frame and the co-addition is performed order-by-order for the calibrated spectra.~For this, we read out the header keyword \emph{HIERARCH ESO SEQ CUMOFF Y} of two subsequent frames $A$ and $B$ and compute the overall nodding offset.~Dividing this offset (in arcseconds) by the scaling factor in the cross-dispersed direction ($0.16\arcsec\,$pix$^{-1}$ for unbinned data) should ideally yield an integer value, as only then can a direct match on pixel-scale level for the overlay of the two frames be achieved and an additional interpolation avoided. The co-added output flux, $F_{AB}$, then becomes 
\begin{equation}
 F_{AB} = F_A + F_{B_s} - (F_B + F_{A_s}), 
\end{equation}
where $F_{A,B}$ are the fluxes in the respective frames and $s$ describes a shift by the nodding offset along the cross-dispersed direction. Since the overlapping part of the output spectrum then contains the sum of two exposures, the obtained counts in that region are divided by two to ensure a proper absolute flux calibration. The same scaling is applied to the propagated errors. 

\subsection{Sky subtraction for point-like objects}
\label{sec:sky_sub}
This section describes the sky spectrum subtraction\footnote{Throughout this paper we refer to the sky spectrum as any spectrum that remains in the slit after subtraction of the primary science target spectrum. This includes any telluric spectrum but, also any other spectrum of an underlying background component.} in UVB and VIS arm spectra. The discussion of the slightly more comprehensive treatment for NIR arm spectra is intended for a forthcoming paper of this series.

\subsubsection{General considerations and removal of artifacts}
In order to compute proper sensitivity functions required for the flux calibration of our dataset, the spectrophotometric standard star signals need to be sky-subtracted.~With{\sc X-shooter}, the standard star observations are usually performed either at the beginning or at the end of the night, often in twilight conditions. Since the default observation mode for these kinds of calibration data is either \emph{OFFSET} or \emph{NOD}, the sky spectrum to be subtracted is usually taken five to ten minutes after the stellar spectrum, depending on the exposure time of the respective standard star. Hence, without proper scaling of the sky level during twilight conditions, the resulting subtraction typically results in a wavelength dependent under- or over-subtraction of up to $15\%$, depending on the twilight conditions. 

The alternative built-in sky subtraction approach of the \xs pipeline (v1.5.0), which defines two sky windows on each side of the science object and models the sky background with two-dimensional {\sc Bezier} splines without any additional sky frame, is better suited for twilight data. However, we found that this method produces unreliable results for some of our tested standard star spectra, in particular for regions where CRH residuals are found or at wavelengths of strong atmospheric absorption or emission.~Since a careful adjustment of the involved fitting parameters did not eliminate the observed sky residuals, we decided not to use any of the pipeline methods. 

To keep the technical complexity manageable, we implemented an order-by-order sky subtraction on the rectified spectra.~For this, two sky window regions of nine pixels in spatial extent are defined at either slit end.~Since residuals of CRHs and bad pixels are likely to compromise the sky signal, a proper treatment is necessary. As v1.5.0 of the \xs pipeline does not propagate CRH positions into the quality control map, we require an algorithm that finds and flags any remaining residuals automatically. For this, at a given slit position, each pixel value is compared to a boxcar median (width: 50 pixels) running along the spectral dispersion direction and a $\kappa\!-\!\sigma$ clipping with $\kappa\!=\!4$ and $\sigma$ comprising the central $68\%$ of the values inside the median box is applied to flag potential outliers. This procedure likewise flags CRHs and sky emission lines in narrow slit setups, requiring an additional step to reliably distinguish between both. In order to do that, we make use of the expected symmetry profile of the emission lines: Each pixel flagged in the first run is compared against the median of the same wavelength bin of the opposite sky window. The same $\kappa\!-\!\sigma$ clipping is applied, however, for this step with $\kappa\!=\!6$. Only if both conditions are fulfilled is the pixel considered a true outlier with respect to the underlying sky signal and its value is replaced with the above described spectral boxcar median.~Both $\kappa$ values have been selected empirically and were tested successfully in wide and narrow slit setups. 

\subsubsection{Spline modeling of the background spectrum}
With all outliers removed from the sky windows, the sky signal needs to be interpolated to the slit positions covered by the star. Fitting each wavelength bin individually along the cross-dispersed direction with a low-order polynomial offers a simple and robust means to handle line emission and absorption accurately once the spectrum is properly rectified. The caveat of this method, however, is its sensitivity to noise, as the spectral correlation length is set to zero by definition. Locally minimizing the residuals inside the sky windows for each bin can therefore lead to an increase in the bin-to-bin scatter outside the sky windows (and within the PSF aperture), as the noise level of each column is considered separately. To circumvent this problem we adopted the above mentioned spline fit approach along the spectral dispersion direction and implemented it to work on rectified spectra. 

\begin{figure}[t]
  \begin{minipage}[b]{\linewidth}
  \centering
  \includegraphics[width=\textwidth]{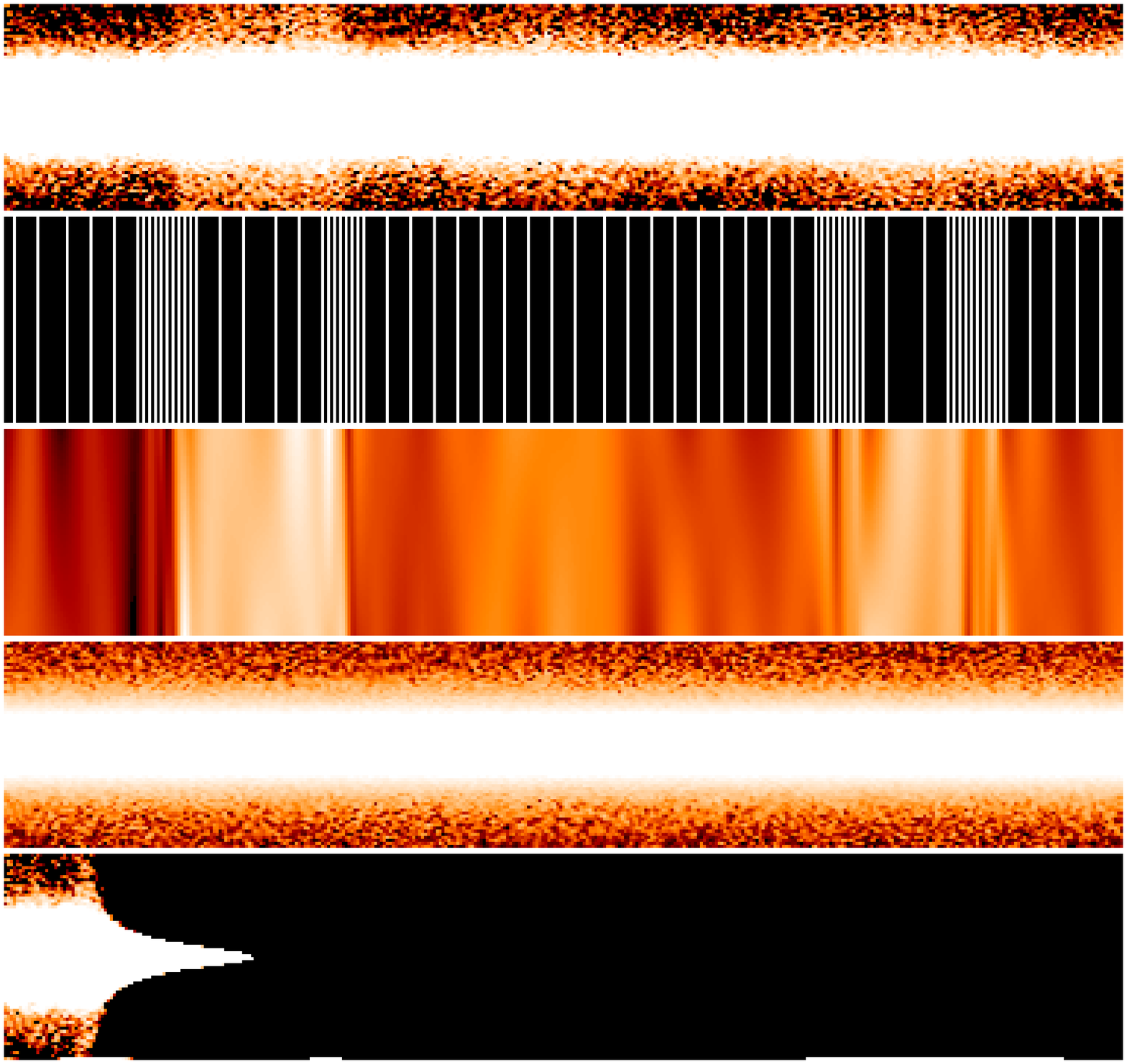}
 \end{minipage}
 \vfill
 \begin{minipage}[b]{\linewidth}
  \centering
  \includegraphics[width=\textwidth, bb= 70 98 613 143,clip]{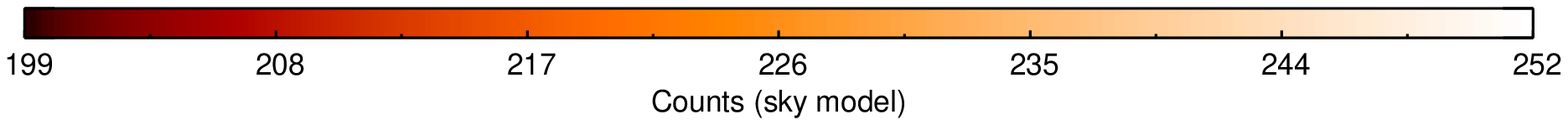}
 \end{minipage}   
 \caption{Sky subtraction sequence for GD71. The top panel shows a rectified spectrum of GD71 with an additional sky component. The break point distribution that forms the basis for the fits to the sky spectrum is shown in the second panel, while the corresponding sky model, based on two separate spline fits at either side of the stellar PSF and an additional interpolation along the cross-dispersed direction, is shown in the third panel. The final sky-subtracted spectrum of GD71 is presented in the fourth panel. For comparison, the fifth panel shows the corresponding result obtained with the \xs pipeline, which failed to produce an acceptable solution for the shown wavelength range ($7742\,\AA$ to $7770\,\AA$) and other parts of the spectrum. The color scaling of the sky model is illustrated in the bar at the bottom.}
\label{fig:sky_sub}
\end{figure}

The benefit of spline fits is their ability to model signals that show small, long-scale (sky continuum convolved with instrumental response) and large, short-scale gradients (line emission, atmospheric absorption windows) at the same time. This offers the ability to efficiently reduce the impact of noise, which is achieved by adjusting the break point density (i.e., denser sampling on strong gradients, sparser sampling on smoother gradients), with which the polynomial pieces are connected, so that the distances between the individual break points reflect the underlying gradient of the sky signal. For this, two step sizes are used: 1) if no features are detected the step width is set to $8$ pixels, and 2) in regions with strong signal gradients (positive and negative) a step width of $2$ pixels is applied, i.e.,~every other pixel is used as break point for the spline fit. As we want to adjust the fitting function to be variable on small scales only where the sky signal itself shows strong variations and, in turn, keep it smooth at all other wavelengths a careful discrimination between actual signal gradients and noise fluctuations has to be made (see Figure~\ref{fig:sky_sub}). 

For this task, we first calculate a variance-weighted average of all pixels contained in the two sky windows along the cross-dispersed direction, resulting in an average sky spectrum (and its standard deviation) which is then further smoothed in spectral dispersion direction by a boxcar average with box size $5$ to reduce bin-to-bin fluctuations.~The resulting spectrum is convolved with a {\sc Laplace}-like kernel of the form $\begin{pmatrix} 1 & 0 &-2 &0 &1\end{pmatrix}$, yielding the second derivative $L$ over a spectral correlation length of five pixels to further suppress pixel-to-pixel variations. The corresponding errors are propagated accordingly (yielding $\sigma_L$).~Zero crossings of the Laplace-convolved spectrum depict large gradient changes if the neighboring values have a point-symmetric shape with respect to the zero crossing with a large absolute value.~For the $i-th$ pixel of the sky spectrum, S, to be flagged as a position with a strong spectral gradient (e.g., transition from continuum to line or vice versa), the surrounding pixel values have to fulfill all three of the following criteria:

\begin{table}[h]
\centering
 \begin{tabular}{ll|l}
    \multicolumn{1}{c}{} & \multicolumn{1}{c}{positive gradient} & \multicolumn{1}{c}{negative gradient}\\
 1. & $\sum\limits_{k=i-5}^{i-3} L_k > \kappa \cdot \sqrt{\sum\limits_{k=i-5}^{i-3} \sigma_{L,k}^2}$ & $\sum\limits_{k=i-5}^{i-3} L_k < -\kappa \cdot \sqrt{\sum\limits_{k=i-5}^{i-3} \sigma_{L,k}^2}$\\
 2. & $\sum\limits_{k=i+2}^{i+4} L_k < -\kappa \cdot \sqrt{\sum\limits_{k=i+2}^{i+4} \sigma_{L,k}^2}$ & $\sum\limits_{k=i+2}^{i+4} L_k > \kappa \cdot \sqrt{\sum\limits_{k=i+2}^{i+4} \sigma_{L,k}^2}$\\
 3. & $\sum\limits_{k=i-5}^{i-3} S_k < \sum\limits_{k=i+2}^{i+4} S_k$ & $\sum\limits_{k=i-5}^{i-3} S_k > \sum\limits_{k=i+2}^{i+4} S_k$\\
  \end{tabular}
\end{table}
The left set of equations has to be fulfilled in case of a positive sky gradient, the right set has to hold if a negative gradient is to be flagged. By comparing averages over the intervals $[i-5, i-3]$ and $[i+2, i+4]$ we ensure that the impact of small-scale variations is degraded with respect to the true gradients inherent in the sky signal.~We carefully adjusted the sensitivity parameter $\kappa$ under the premise that even very faint lines or blended features are accurately flagged.~As a side effect, the requirement to include very faint lines leads to a significant flagging of noise spikes, thereby potentially lowering the spectral correlation length of the sky model at wavelengths without features and, hence, resulting in a slightly noisier fit. After detailed experimentation, we found $\kappa\!=\!0.3$ to work well even for long exposed twilight data with many emission and absorption features in the red part of the VIS arm. If the S/N of the average sky spectrum drops below one, then we set $\kappa\!=\!0.8$, which further minimizes the impact of noise, as no spectral features are expected in such a low S/N environment. In general, the above mentioned criteria have been carefully tested and optimized for many different observational setups and S/N scenarios and should therefore be widely applicable.

With all sky features flagged, seven break points with a stepping of two pixels are symmetrically distributed around each flagged position, while checking and correcting for possible overlaps within this distribution.~Each interspace between the flagged positions is then filled up with break points with a stepping of eight pixels if the median number of counts within this space is $>\!30$ and a stepping of $20$ pixels if the median number of counts is $<\!30$.~Increasing the step size when only a few sky photons are detected enforces a smoother sky model and makes the approach more robust in low S/N scenarios.~The threshold number has been empirically determined and performs well for all tested observations.~Overall, this implementation yields a break point distribution whose spacing is adjusted to the spectral gradient of the sky signal (see Figure~\ref{fig:sky_sub}).

For the actual sky model, we take the weighted average for both sky windows (along the cross-dispersed direction) at every wavelength sampling point, as computed before, and then fit the obtained sky spectrum in the lower and upper sky window independently along the spectral direction.~To further suppress noise artifacts, we additionally smooth the sky signal (before fitting) along the spectral direction with a boxcar median (21 pixels box size) inside those wavelength ranges where the break point stepping is 20 pixels.~Since an accurate error propagation for spline fits can be cumbersome \citep{splineerror}, we implemented a Monte Carlo approach, which simulates and fits each sky window $200$ times with cubic splines based on the break point distribution,~accounting for the noise level inside the sky windows.~To decrease the impact of individual pixel values within the computed break point sample, we shift the global break point distribution by half the distance between each break point pair and repeat the Monte Carlo simulation. The final solution is the average (including its standard deviation) of the 400 fit realizations.~Based on the obtained mean values and standard deviations the two models on either side of the stellar PSF are then linearly interpolated to all slit positions for each wavelength bin. This interpolation is again implemented as a Monte Carlo simulation, with $200$ random realizations accounting for the $1\,\sigma$ errors of the spline fits, yielding a two-dimensional sky model with consistently estimated uncertainties that can be propagated into the total error budget.

Allowing for a gradient during the linear interpolation along the cross-dispersed direction is necessary for two independent reasons: 1) if the spectrum is obtained during twilight the sky can show intrinsic gradients, even along the relatively short slit length of 11\arcsec\ and 2) residuals from the inter-order scattered-light subtraction (see Section~\ref{sec:io_background_model}) might show up as an additional gradient component, which can then be removed by subtracting a properly adjusted sky surface.~To illustrate this sequence of steps, Figure~\ref{fig:sky_sub} shows a part of the VIS spectrum for star GD71 before and after sky subtraction, together with the used break points and the obtained sky model. 

To provide the user with additional quality control possibilities, the software outputs the break point mask, the sky model, and a sky subtraction residual image as separate \emph{FITS} files so that any fitting parameter can be adjusted conveniently if desired.

\subsection{Optimal extraction of point-like objects}
\label{sec:opt_extr}
Extracting a one-dimensional spectrum out of two-dimensional data is a complex but mandatory task for all point-like objects \citep{horne86, cushing04}. Although the globular clusters in our dataset are well resolved with \xs, the sensitivity functions required for absolute flux measurements rely on accurate observations of point-like standard stars\footnote{The presented extraction approach was designed to be applied to our spectrophotometric standard star observations.~However, there is no limitation on the applicability to any other point-like object.}.

The goal of our optimal extraction technique is to balance the requirements of limiting the propagation of background noise on the one hand, and the extraction of the maximum amount of information on the stellar SED on the other hand. Hence, the cross-dispersed PSF needs to be properly modeled so that the threshold between signal and noise can be accurately estimated. 

Our optimal extraction routine integrates the counts and their associated errors (from the error map) of each wavelength bin over an extraction aperture that maximizes its integrated S/N and determines the associated flux losses in the wings from the model PSF. Because of the principal lack of through-slit images, additional slit losses (occurring when the light rays pass through the slit) can only be estimated from the one-dimensional cross-dispersed signal that is imaged onto the detector. 

This technique is, in principle, similar to the optimal extraction routine of \cite{horne86} which is designed to iteratively find the best source profile fit along the spectral direction from data affected by noise and cosmic ray hits.~Our approach is insofar different as we model the source profile along the cross-dispersed direction and propagate the full variance of the data {\it and} the uncertainty of the profile fit, while the \citeauthor{horne86} technique does not account for the r.m.s.~of its profile fit in each iteration. With our prior knowledge of the full spatial light distribution along the slit for the entire wavelength range and its uncertainty propagated into the variance frame (including cosmic ray hits and CCD cosmetics, see Sects.~\ref{sec:io_sub} and \ref{sec:bp}), we can {\it directly} determine the optimal extraction aperture which maximizes the S/N ratio of the extracted spectrum. In the following, we explain this in more detail.

\subsubsection{Modeling the cross-dispersed PSF}
\begin{figure}[t]
\centering
\includegraphics[width=\linewidth]{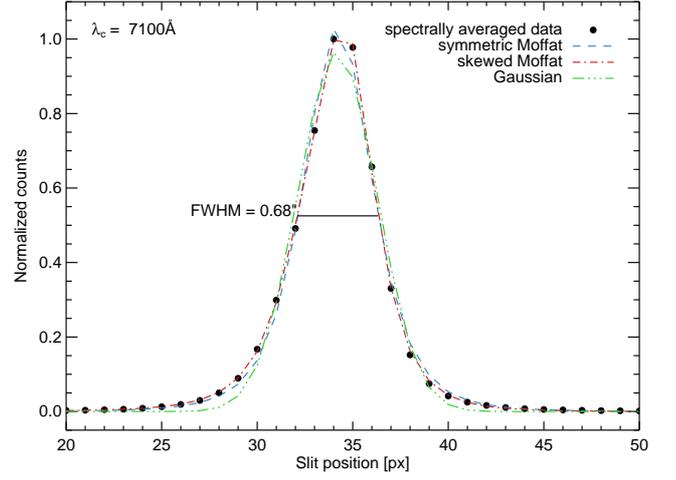}
\caption{Cross-dispersed \xs PSF and its best-fit models for the flux standard star BD+174708 in echelle order eight of the VIS arm ($\lambda_c = 7100 \,\AA$). The fit to the spectrally averaged data (black circles) is considerably improved if a skewness term is added to the Moffat profile (dash-dotted red curve) as compared to its symmetric counterpart (dashed blue curve). The measured FWHM is $4.22\,{\rm pix}\!=\!0.68\arcsec$, whereas the DIMM seeing is indicated as $0.52\arcsec$. The best-fit Gaussian is overplotted with a green dash-dotted line for comparison.} 
\label{fig:moffat_psf}
\end{figure}

In order to obtain crude, but robust estimates for the cross-dispersed PSF in each echelle order even for low S/N data, we spectrally average over $10\,\AA$ around the blaze wavelength of each order and normalize the peak of the count level to unity.~A skewed Moffat profile \citep{moffat} is fitted with a {\sc Levenberg-Marquardt} least-squares method (IDL implementation by \citealt{idlmpfit}) to the resulting one-dimensional PSF.~The Moffat profile is preferred over a Gaussian shape as the \xs cross-dispersed PSF shows relatively pronounced wings, which cannot be reproduced with Gaussian.~The exact parametrization is of the form
\begin{equation}
 {\rm PSF} = \frac{ a_1  }{ (u^2 + 1)^{a_4}  }(1 + \mathcal{P}(a_5 u + a_6 u^2 + a_7 u^3)),
 \label{eqn:moffat}
\end{equation}
with $u\!=\!(y\!-\!a_2)\,a_3^{-1}$ and $y$ being the slit coordinate. The asymmetry term, $\mathcal{P}(\dots)$, becomes necessary especially for good seeing conditions, when the optical distortions of the instrument become significant.~Figure~\ref{fig:moffat_psf} shows such a scenario for one observation of the flux standard BD+174708, with a measured FHWM of $0.68\arcsec$ at a central wavelength of $\lambda_c\!=\!7100\,\AA$, where both the symmetric and skewed best-fit models are overplotted. In general, we find significant symmetry deviations for seeing values corresponding to a FWHM $\lesssim\!0.8\arcsec$.~The penalty factor $\mathcal{P} \in [0,1]$ weights the expansion term and is introduced for low S/N data, because the additional asymmetry parameters might lead to unconstrained fitting parameters and divergent PSF models. In the first fitting run $\mathcal{P}$ is fixed at $1$ and only orders with an averaged ${\rm S/N}\!>\!100$ are fitted and a best-fit PSF model is obtained. 

 The resulting model parameters at the blaze wavelengths of each order with a ${\rm S/N}\!>\!100$ serve as initial guesses for additional fits at other wavelength ranges.~The necessity for multiple PSF models within one echelle order arises from the wavelength-dependent seeing and the instrumental distortions that show commensurable variations on a $3\%$ level within each order. In total, each order is sampled at ten equidistant spectral positions, using a boxcar average along the spectral direction to suppress noise. Averaging the data in such boxes with typical sizes of $100-400$ pixels is complex, since, in case of a broken atmospheric dispersion correction (ADC) unit and/or poor wavelength calibration, the PSF centroid position is a function of wavelength, with typical shifts of $1\!-\!2$ pixels from one order edge to the other\footnote{For proper wavelength calibrations, the centroid drift is typically stable at a $0.1$ pixel level, which corresponds to the internal accuracy of the rectification process.}. A spectral average of such a drifting PSF potentially leads to an artificial widening of the PSF. Therefore, we first trace the centroid position along each order by boxcar averaging over ten adjacent wavelength bins (i.e., with negligible centroid shifts within such a box) and fit the resulting spatial profile with a symmetric Moffat function of fixed width. This setup is chosen for reasons of execution speed, technical simplicity and robustness, and does not affect the accuracy of the derived information content. The symmetric profile (first term of Equation~\ref{eqn:moffat}) is preferred for this task, as the parameters of its asymmetric counterpart have been shown to exhibit degeneracies and, hence, discontinuities can be observed if only a single parameter (centroid) is considered.
 
Once the centroid drifting function is known, the data are copied to a new array and by linear interpolation relocated to a new grid that is centered around the above fitted centroid trace. As this grid is now free from centroid shifts, the necessary boxcar averages can be computed without any further considerations, yielding a dataset with ten PSFs at equidistant spectral positions per order. If the integrated signal-to-noise $S_i$ is below 20, no fit to the data is performed at all and the model PSF is taken from the next sampling point where a model could be properly estimated. For sampling points with ${S_i}\!>\!20$, $\mathcal{P}$ in Equation (\ref{eqn:moffat}) is adjusted to the quality of the data by comparing it to a threshold signal-to-noise ${S}_t$ above which the data are typically good enough to perform an unpenalized ($\mathcal{P}\!=\!1$), i.e.,~asymmetric fit.
 
First, all parameters in Equation~(\ref{eqn:moffat}) are fitted simultaneously at fixed $\mathcal{P}\!=\!1$, whereas in a second run, if ${S}_i\!<\!{S}_t$, the asymmetry parameters are kept fixed at the values of the first fit, the penalty factor is set to $\mathcal{P}\!=\!({S}_i/{S}_t)^2$, and only the symmetric Moffat profile parameters are updated. With the choice of ${S}_t=100$ we find that this method produces robust results for both low S/N data, where the actual shape of the PSF is not well constrained and the symmetric Moffat parameters is the only information that can be reliably extracted, as well as for high S/N data, where significant deviations from symmetry can be accurately modeled. In cases where all boxcar averages show ${\rm S/N}\!<\!20$, the PSF is modeled with the penalized approach at the sampling point with the highest S/N, and subsequently copied to all other wavelength positions.

The cross-dispersed range that is used for the PSF fit depends on the mode of observation. For \emph{STARE} data, all pixels are included in the fit, whereas in \emph{NOD} mode, the negative ghost PSFs at the slit edges need to be excluded, as they would otherwise affect the quality of the fit in the wings of the central PSF. Our truncation limits on the fitting region are based on two estimates: 1) the optimal extraction aperture (within which the integrated S/N is maximized) and 2) the zero-crossings of the averaged data (transition between central, positive PSF and negative PSFs at the slit edges).~We take the mean value of both and round the result to the nearest integer. This choice performs reliably well, even in cases of bad seeing ($>\!2\arcsec$) and small nodding offsets where the wings of the central PSF are substantially affected by their negative counterparts. As a further stability constraint for the convergence of the fit we impose additional boundary conditions and set the outermost four pixels at both slit ends to zero for \emph{STARE} data, while this margin is increased to 15 pixels for \emph{NOD} mode observations, for which the combined cross-dispersed dimension of the slit is significantly larger.

\begin{figure}[t]
\centering
\includegraphics[width=\linewidth]{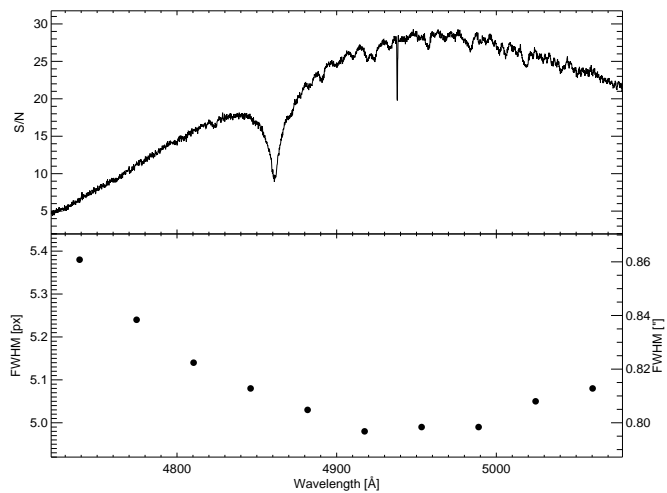}
\caption{Optimally-extracted S/N spectrum (top panel) and measured FWHM (bottom panel) in echelle order 10 (UVB arm) for HD 38237. The downward S/N spike at $4938\,\AA$ arises from a cold spot on the CCD and demonstrates the quality of our optimal extraction procedure. The FWHM shows variations on a five percent level with a minimum close to the blaze wavelength.} 
\label{fig:sn_fwhm}
\end{figure}

In order to increase the numerical accuracy when estimating the flux losses accompanying the optimal extraction, the obtained PSF models are computed on a grid that is oversampled ten times with respect to the original cross-dispersed resolution and subsequently propagated to all other wavelengths by fitting a second-order one-dimensional polynomial along the spectral direction for all slit positions. The choice of the polynomial order is supported by the smooth FWHM vs.~$\lambda$ trend, as shown in the bottom panel of Figure~\ref{fig:sn_fwhm}, which typically shows a minimum around the blaze wavelength. The resulting wavelength-dependent PSF model contains all the necessary information on the PSF shape variability and is then back-transformed to the coordinate system of the original data (i.e.,~reintroducing the PSF's centroid drift) in order to be used for the object extraction.

\subsubsection{Target signal extraction and flux loss determination}
\begin{figure}[t]
\centering
\includegraphics[width=\linewidth]{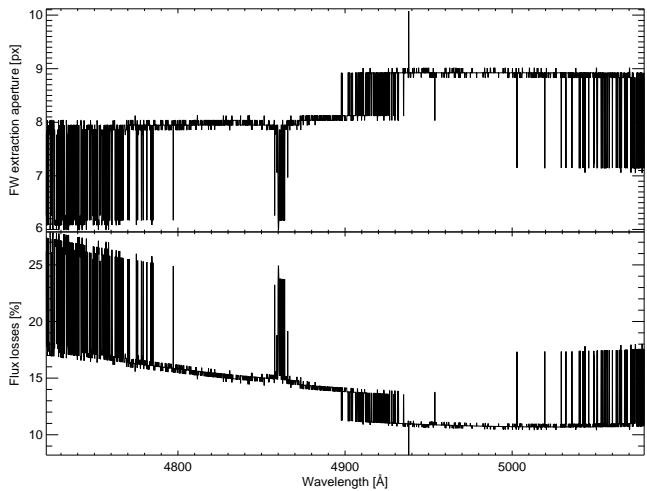}
\caption{Full width optimal extraction apertures (top panel) and their associated flux losses (bottom panel) in echelle order 10 (UVB) for the star HD 38237. The extraction apertures resemble a discrete step function, as the S/N only changes at the transition between adjacent pixels. The corresponding flux losses follow this step function, but show an underlying smooth component that arises from the continuous PSF shape variation within the echelle order (see bottom panel of Figure~\ref{fig:sn_fwhm}).} 
\label{fig:extr_ap_fl_loss}
\end{figure}

The optimal extraction is performed by integrating each data bin along the slit dimension to its maximum S/N aperture, and correcting for the clipped counts by extracting the flux fraction of the truncated regions from the corresponding analytic model.~The same technique is applied to the respective error frame, where the uncertainties of the clipped counts are estimated with a Poisson noise model, which we have found to be a conservative upper limit to the error of the profile fitting process as estimated by a dedicated Monte-Carlo simulation based on the covariance matrix of the model fit parameters.~Figure~\ref{fig:extr_ap_fl_loss} shows the computed extraction apertures and their corresponding flux losses for the star HD 38237 (spectral type A3) in echelle order 10 (UVB). While the extraction aperture function (top panel) shows discrete steps in integers of full pixels in addition to a fluctuation that is related to the combination of a non-centered centroid within a pixel and a symmetric extraction aperture around that centroid, the associated flux losses (bottom panel) pick up these discontinuities, but show an additional parabolic shape that is caused by the above mentioned wavelength dependence of the PSF. The amplitude of this systematic effect is three percent.

Estimating the flux losses due to the finite slit aperture requires an additional two-dimensional modeling of the PSF. As \xs does not offer the possibility of acquiring through-slit images, we estimate the unaccessible second dimension from the cross-dispersed profile.~For this, we use a {\sc Levenberg-Marquardt} least-squares fit to construct a symmetric two-dimensional PSF that, if collapsed along the spectral dimension, fits the measured cross-dispersed data.~This is superior to the \cite{horne86} approach which does not provide a parametric solution of the cross-dispersed profile that could be used as a basis for the subsequent two-dimensional PSF model.~Spherical symmetry is chosen as a prior, since we do not have any information about the potential occurrence of pre-slit skewness along the direction that becomes the spectral direction inside the spectrograph. The slit losses can then be estimated by setting an aperture resembling the long-slit around the PSF model. The aperture size is computed based on the information about the slit width from the file header. The slit loss estimation is performed for each order individually at the wavelengths shown in Col. 3 of Table \ref{tbl:sensfunc} and the obtained values are then fitted globally with a third-order polynomial to obtain a smooth function for the entire spectral range of the respective instrument arm. The fitting weights are based on the corresponding S/N values.

\subsubsection{Validation of extraction quality}
\begin{figure}[t]
\centering
\includegraphics[width=\linewidth]{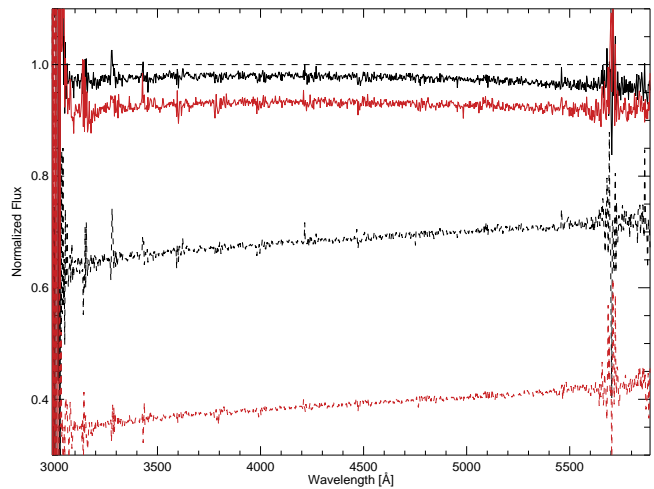}
\caption{Normalized fluxes of GD71 in the UVB arm obtained with two different slit widths and two different extraction methods. The solid black (red) curve shows the optimally extracted and flux loss corrected data of the $1.0\arcsec$ ($0.5\arcsec$) slit exposure, whereas the corresponding dashed curves show the conventional approach of collapsing the entire slit. All exposures have been normalized to the detected counts of a $5.0\arcsec$ slit exposure, for which no significant flux losses have been measured. The data shown here are the merged result of all UVB orders. We note that the curves have been smoothed so that systematic trends can emerge.} 
\label{fig:GD_71_slit_loss_correction}
\end{figure}

Optimally extracting the two-dimensional spectrum and correcting for both optimal extraction aperture flux losses and additional slit losses is mandatory if an accurate flux calibrated one-dimensional spectrum with a minimal noise component is required.~To validate our approach and measure its accuracy, we reduced and compared a sequence of three observations of the same photometric standard star GD71 ($\rm{m_V}\!=\!13)$ that were taken with similar exposure times (100 seconds) but different slit widths of $5.0\arcsec$, $1.0\arcsec$, and $0.5\arcsec$. With a typical seeing of $1.0\arcsec$ in all three measurements (UVB), we defined the $5.0\arcsec$ slit size observation to be the reference case with negligible slit losses, and normalized the counts of the other two observations relative to this reference spectrum. Figure \ref{fig:GD_71_slit_loss_correction} shows the resulting flux ratios for both cases, where each observation has been optimally extracted (solid lines) and the individual echelle orders merged into one spectrum (for details on the merging process, see Section~\ref{sec:fluxcal}).~For comparison, the conventional approach of collapsing the entire slit is overplotted as well (dashed lines). For the $1.0\arcsec$ slit width (black dashed line), only about $65\%$ of the overall flux at $4000\,\AA$ reaches the detector, whereas this number increases to approximately $72\%$ at the red limit of the UVB arm ($5850\,\AA$). This trend correlates with the wavelength-dependent seeing, which is $1.05\arcsec$ at the blue end and $0.75\arcsec$ at the red end, according to the on-detector FWHM. When correcting for slit losses, the ratio reaches $98\%$ between $3700\,\AA$ and $4200\,\AA$, and slightly drops to $97\%$ at the blue end and to $96\%$ at the red end (black solid line). We believe that this residual curvature of the corrected flux may be caused by the ADC unit that almost perfectly corrects the PSF centroid position for wavelengths close to the zero-deviation wavelength ($4050\,\AA$ for the UVB arm, see \citealt{vernet2011}), but shows minor residuals for all other wavelengths. This scenario is supported by the orientation of the slit tilt angle, which was set to $\sim\!25^{\rm o}$ relative to the parallactic angle for the presented observation, implying that any ADC imperfections might potentially introduce flux losses at the slit entrance.~In addition to the smooth, global deviations from 100\%, we find small systematic variations of $\lesssim\!2\!-\!3\%$ between the blue end (typical S/N~$\lesssim\!10$) and the center (S/N~$\gtrsim\!20$) for some of the orders with wavelengths $>\!3500\,\AA$, which imposes a S/N dependent limit on the absolute flux calibration as discussed in Section~\ref{sec:fluxcal}. 

Decreasing the slit size to 0.5\arcsec\ leads to a similar scenario; however, only $\sim\!40\%$ of the flux is captured by the detector (red dashed line in Figure~\ref{fig:GD_71_slit_loss_correction}).~Applying the slit corrections lifts the ratio to $0.94$ at the zero-deviation wavelength, with the same drop-offs at both sides as in the $1.0\arcsec$ slit-width case (red solid line). For both the $0.5\arcsec$ and $1.0\arcsec$ slit widths, a slightly off-centered PSF within the slit caused by the limits of the VLT's pointing accuracy could be responsible for the observed systematic underestimation. However, this is purely speculative and cannot be examined in greater detail with the current data at hand.

\begin{figure}[t]
\centering
\includegraphics[width=\linewidth]{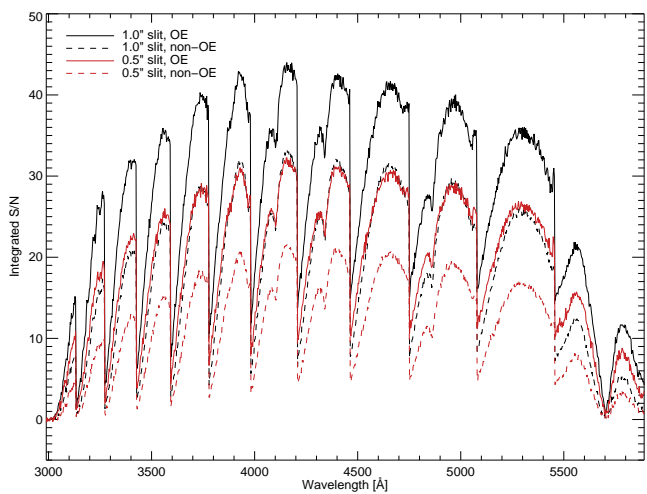}
\caption{S/N curves of GD71 in the UVB arm obtained with two different slit widths and two different extraction methods (100 seconds exposure time). The solid black (red) curve shows the S/N of the optimally extracted $1.0\arcsec$ ($0.5\arcsec$) slit-width data and the corresponding dashed curves show the conventional approach of collapsing the entire slit. We note that the curves have been smoothed so that systematic trends can emerge.} 
\label{fig:GD71_slit_sn}
\end{figure}

The discontinuous peaks with amplitudes of $5\!-\!10\%$ bluewards of $3500\,\AA$ have been investigated carefully, but so far no satisfying explanation can be given. The spectral positions correspond to wavelengths right next to an order overlap, for which the blue part of the subsequent echelle order constitutes the only component of the merged spectrum. After extensive testing, we tend to rule out any residuals from additive effects, i.e.,~inter-order background subtraction or sky subtraction.~We noticed a peculiar asymmetry in the PSF for the two narrow slit observations at these wavelengths, which apparently is not entirely modeled by our PSF parametrization. More accurate examinations proved to be unfruitful, as the integrated S/N per wavelength bin is $\lesssim\!5$ at the affected wavelengths (see Figure~\ref{fig:GD71_slit_sn}).~Furthermore, we noticed that the PSF shape at $5700\,\AA$ considerably opens up and returns back to normal size within a short spectral range ($\lesssim\!100\,\AA$).~Since this wavelength range coincides with the spectral position of the transparency dip of the UVB/VIS dichroic element (see Section~\ref{sec:fluxcal} for further details), we suspect that the PSF change happens before the slit (and outside the actual UVB/VIS spectrographs). A dedicated high S/N calibration sequence with various slit sizes in different observational scenarios (slit tilt angle, airmass, seeing) would be required to better quantify the instrumental PSF and its dependence on environmental parameters. 

The wavelength dependent increase in S/N of our optimal extraction method is shown in Figure \ref{fig:GD71_slit_sn}. In the $1.0\arcsec$ slit-width scenario, the median improvement is 1.51 (arithmetic mean 1.70), although it can be significantly higher in the blue part of the spectrum and at the order edges, where the overall instrumental response drops. For the $0.5\arcsec$ slit, the median increase is 1.75 (arithmetic mean 1.93). 

We note that the structure of the optimal extraction technique presented here can also be used to extract resolved or marginally resolved objects.~In these cases, the Moffat light profile then needs to be replaced with a robust parametrization of the object's cross-dispersed light profile.

\subsection{Absolute flux calibration and order merging}
\label{sec:fluxcal}
To flux-calibrate our UVB and VIS arm \xs data, we compute the sensitivity functions based on optimally extracted spectrophotometric standard star spectra for each order individually. For this, we follow the scheme used by the \emph{IRAF} tasks \emph{standard}, \emph{sensfunc}, and \emph{calibrate}. 

\subsubsection{Method}

\begin{figure*}[t]
\centering
\includegraphics[width=\linewidth]{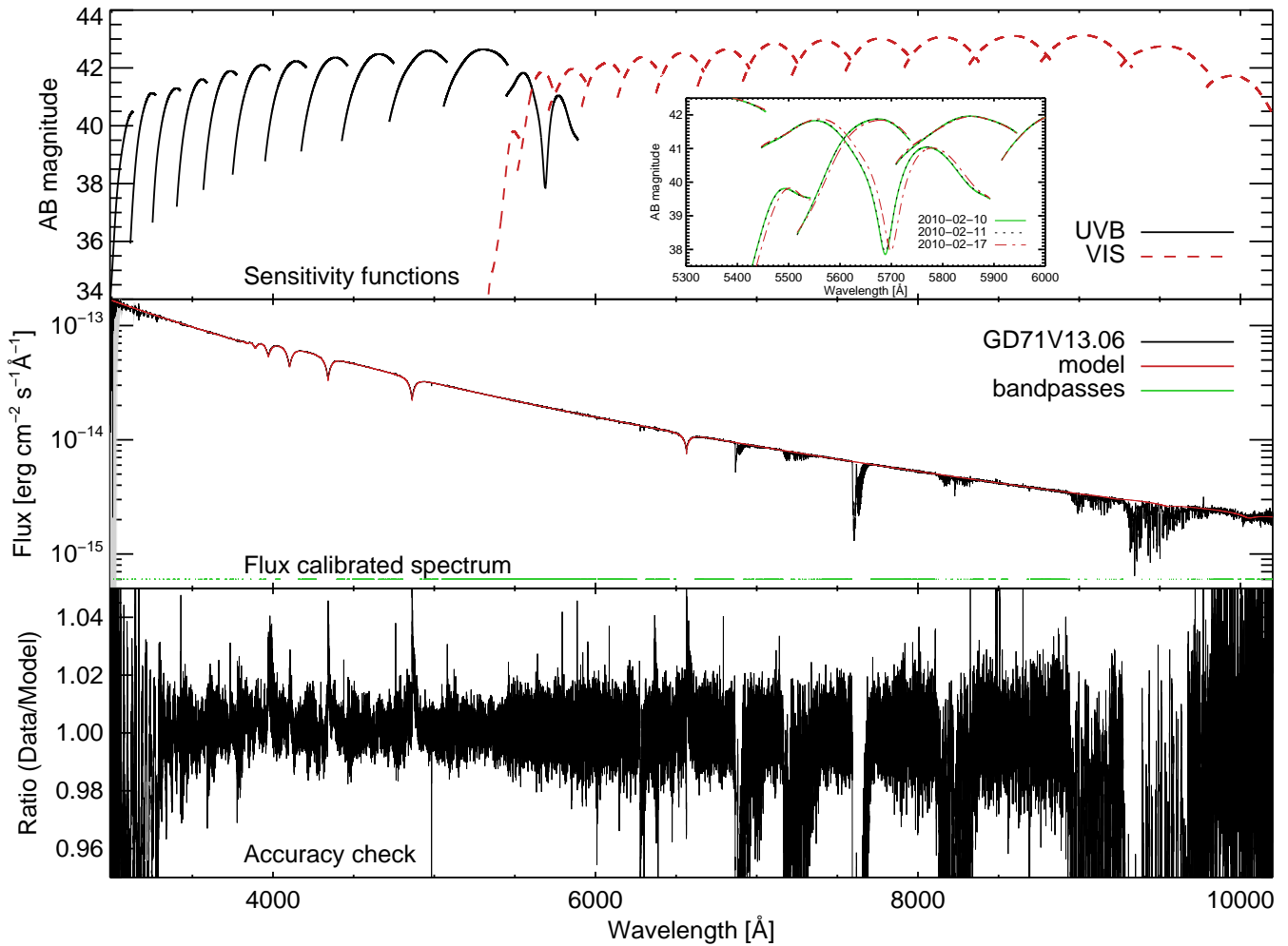}
\caption{UVB and VIS sensitivity functions and the resulting flux-calibrated spectrum for GD71. The top panel shows the order-by-order sensitivity functions in AB magnitudes for both the UVB (solid black line) and VIS (dashed red line) arm. The temporal variation of the instrumental throughput at the transition regime of the UVB/VIS dichroic is illustrated in the inset, where we show three different sets of sensitivity functions, obtained on three different nights in the relevant wavelength range. There, the solid green curve indicates our reference set (obtained on 2010 February 10), the dotted black line the set obtained 24 hours later, and the dot-dashed red curve depicts the sensitivity functions one week later. The middle panel shows the final flux-calibrated output for GD71 (solid black line), where all echelle orders of both instrumental arms have been merged into one spectrum. The underlying model SED is overplotted in red. The propagated observational $1\sigma$ uncertainties are indicated with a smoothed, symmetrical gray band around the mean values and are hardly visible for most of the spectral range. The bandpass selection used as input for the count integration is indicated in green at the bottom of the panel. Atmospheric absorption bands have not been corrected for. The ratio between model and data is shown in the bottom panel. Systematic deviations are typically $\lesssim 0.5\%$, except for wavelength ranges covered by stellar features, where they can increase to $\lesssim 4\%$. We note that the flux-calibrated spectrum of GD71 is plotted with a very high resolution and so the electronic version of this plot is highly zoomable.}
\label{fig:sens_func}
\end{figure*}

In a first step, bandpasses that homogeneously sample the continuum of the respective standard star SED are defined.~We carefully mask both stellar and telluric spectral features and use a generic bandpass width of $5\,\AA$ with a sampling step size of $5\,\AA$ if the continuum is well defined.~As some echelle orders (e.g., ~VIS 14) are almost completely affected by atmospheric absorption, this task proved to be laborious and forced us to use significantly smaller and irregularly distributed bandpasses for certain wavelength ranges. This task was performed for both our standard stars (GD71V13.06, BD+174708) in both instrument arms (UVB, VIS) and the resulting bandpass samples are generally valid for these stars and can be used for any observation of that kind. Our bandpass selection for GD71V13.06 is indicated as a green baseline in the middle panel of Figure~\ref{fig:sens_func}. 

With the bandpasses defined, the standard star spectrum is integrated over all bandpasses and the resulting counts are stored in tabulated form together with the information on the bandpass location (central wavelength), bandpass width, and corresponding flux (at central wavelength) of a given model SED. For the latter, we use the pipeline-packaged models that are based on observations by \citet{hamuyspecphot} and \citet{vernetspecphot}. They cover the entire wavelength from $3000\,\AA$ to $25000\,\AA$ at varying spectral resolutions; however, they need to be interpolated to a finer sampling to match the dispersion of our data.~These tables are used as input for \emph{IRAF sensfunc}, where separate sensitivity functions for each echelle order are computed including corrections for the effective airmass. As response fitting functions, we generally use cubic splines or {\sc Chebyshev} polynomials, depending on the bandpass sampling of the respective order. The fit is manually cleaned from any outliers caused by bad pixel or CRH residuals and the function type and polynomial order are adjusted interactively if required.~This process is performed iteratively with visual verification of convergence, as orders only weakly constrained by the bandpass selection react very sensitively to a change in the fitting parameters.~We note that first correcting the science frame by the blaze function as derived from the flat lamp frame did not improve the quality of the final sensitivity function fit solution.~By contrast, such a procedure would introduce additional complications, such as line emission in the flat lamp spectrum and a variation of the lamp's SED shape due to temporal voltage/temperature changes (see Section~\ref{sec:ff}).

The subsequent merging process of the flux calibrated orders is performed with an error weighted average in the order-overlap regions and the errors are propagated accordingly. The same merging technique is also applied when the flux-calibrated results of the two instrument arms are to be merged into one global spectrum. 

\subsubsection{Sensitivity function evaluation}
Table \ref{tbl:sensfunc} shows an overview of relevant calibration parameters for our reference GD71V13.06 observations. The top panel of Figure \ref{fig:sens_func} shows the resulting sensitivity functions. The functions are generally of parabolic shape, except for order 12 of the UVB arm. Instead of showing a typical response maximum at the blaze wavelength, this particular order features a pronounced dip close to the central wavelength of the echelle order. This effect is caused by a non-monotonous drop of the dichroic transparency with local side maxima around $5700\,\AA$ \citep[Figure 4 therein]{vernet2011} and goes along with a PSF broadening in the cross-dispersed direction (see Section~\ref{sec:opt_extr}). When comparing observations with different time stamps, the dip positions and, hence, the associated sensitivity functions, turn out to be temporally variable, as the local maxima of the dichroic's transparency function are shifted along the dispersion axis (see inset in the top panel of Figure~\ref{fig:sens_func}).~While the instrumental throughput remained roughly constant within a time baseline of 24 hours in the shown case (solid green and dashed black curve), it varied by $20\!-\!30\%$ within seven days (dot-dashed red curve). For other cases, we find a similarly high variation amplitude even within a single night. The true nature of this effect has yet to be unveiled and its impact quantified. Nevertheless, checking the ESO Ambient Condition Database\footnote{http://archive.eso.org/asm/ambient-server} we speculate that ambient humidity might be responsible for the observed instrument behavior. On the respective nights, the humidity changed by a factor of three between the observation on 2010 February 17 and the ones on 2010 February 10 and 11, while the temperature remained constant within $2^\circ$C.

Comparing the sensitivities of the UVB and VIS arm we discovered that, although they are very similar in shape, the fitted functions are of comparable amplitude at the overlap regions of the individual orders for the VIS arm, whereas pronounced discontinuities between adjacent orders can be observed in the UVB arm. This effect originates from the fact that the red part of each UVB echelle order is systematically trimmed by the order extraction mechanism of the ESO pipeline by $\sim 80\,\AA$ (using the {\sc Spectral Format Table} provided with the pipeline kit). Unfortunately, the cropped data typically have a higher S/N than the corresponding ones at the same wavelengths in the subsequent order, resulting in much more pronounced steps in the merged spectrum compared to the VIS arm, where signals of comparable quality overlap. The reason for the differences in the order extraction between the two instrument arms is unclear and so this issue has to remain unsolved.

\begin{table*}
 \begin{tabular}{ p{0.7cm} p{0.9cm} p{1.8cm} p{1.9cm} p{1.8cm} p{1.8cm} p{2.0cm} p{1cm} p{1cm} p{2cm} cccccccccc }
 \hline
 \hline
  \textbf{Order} & \textbf{Spectral order} & \textbf{Min. wavelength [$\,\AA$]}& \textbf{Slit loss model wavelength [$\,\AA$]}& \textbf{Max. wavelength [$\,\AA$]} & \textbf{Avg. dispersion [$\,\AA$/px]} & \textbf{Fitting function / polynomial order} & \textbf{No. of bandpasses} & \textbf{RMS [$10^{-2}$]} & \textbf{Remarks}\\
  \hline
  \multicolumn{9}{c}{\textbf{UVB}}\\
  \hline
  1 & 24 & 2989 & 3125 & 3134 & 0.097 & Spline3/4 & 8 & 0.44 & atmospheric absorption, low S/N\\
  2 & 23 & 3115 & 3260 & 3273 & 0.103 & Spline3/4 & 10 & 0.50 & atmospheric absorption\\
  3 & 22 & 3253 & 3400 & 3425 & 0.109 & Spline3/4 & 10 & 0.03 & atmospheric absorption\\
  4 & 21 & 3403 & 3550 & 3592 & 0.115 & Chebyshev/8 & 9 & 0.02 & \\
  5 & 20 & 3568 & 3680 & 3777 & 0.122 & Chebyshev/8 & 21 & 0.17 & \\
  6 & 19 & 3750 & 3920 & 3981 & 0.129 & Spline3/4 & 9 & 0.07 & $H_\epsilon, H_\zeta, H_\eta$\\
  7 & 18 & 3951 & 4150 & 4208 & 0.137 & Chebyshev/6 & 17 & 0.10 & $H_\delta$\\
  8 & 17 & 4175 & 4370 & 4463 & 0.146 & Chebyshev/8 & 28 & 0.18 & $H_\gamma$\\  
  9 & 16 & 4427 & 4666 & 4751 & 0.155 & Chebyshev/10 & 51 & 0.21 & \\
  10 & 15 & 4721 & 4970 & 5079 & 0.166 & Chebyshev/12 & 44 & 0.14 & $H_\beta$\\
  11 & 14 & 5058 & 5300 & 5456 & 0.178 & Chebyshev/15 & 80 & 0.19 & \\
  12\_1 & 13 & 5446 & 5560 & 5693 & 0.196 & Spline3/15 & 49 & 0.20 & bluewards of dichroic dip\\
  12\_2 & 13 & 5693 & 5770 & 5893 & 0.184 & Spline3/15 & 40 & 0.25 & redwards of dichroic dip\\
  \hline
    \multicolumn{9}{c}{\textbf{VIS}}\\
  \hline
  1 & 30 & 5337 & 5480 & 5543 & 0.103 & Spline3/8 & 36 & 1.85 & dichroic variability, low S/N\\
  2 & 29 & 5516 & 5640 & 5737 & 0.106 & Spline3/8 & 40 & 0.15 & dichroic variability\\
  3 & 28 & 5709 & 5840 & 5945 & 0.111 & Spline3/8 & 46 & 0.20 & dichroic variability\\
  4 & 27 & 5915 & 6050 & 6169 & 0.115 & Spline3/8 & 46 & 0.15 & \\
  5 & 26 & 6136 & 6250 & 6411 & 0.120 & Spline3/9 & 40 & 0.16 & \\
  6 & 25 & 6375 & 6525 & 6672 & 0.126 & Spline3/6 & 32 & 0.12 & $H_\alpha$\\
  7 & 24 & 6633 & 6800 & 6955 & 0.132 & Chebyshev/9 & 48 & 0.22 & atmospheric absorption window\\
  8 & 23 & 6913 & 7100 & 7263 & 0.138 & Chebyshev/9 & 53 & 0.30 & atmospheric absorption window\\  
  9 & 22 & 7218 & 7420 & 7600 & 0.145 & Spline3/6 & 53 & 0.18 & atmospheric absorption window\\
  10 & 21 & 7550 & 7770 & 7970 & 0.152 & Spline3/8 & 60 & 0.28 & atmospheric absorption window\\
  11 & 20 & 7915 & 8150 & 8378 & 0.160 & Spline3/6 & 50 & 0.27 & atmospheric absorption window\\
  12 & 19 & 8317 & 8600 & 8830 & 0.169 & Spline3/6 & 61 & 0.26 & \\
  13 & 18 & 8762 & 9050 & 9329 & 0.179 & Chebyshev/6 & 47 & 0.51 & atmospheric absorption window\\
  14 & 17 & 9258 & 9550 & 9810 & 0.193 & Chebyshev/6 & 24 & 0.92 & atmospheric absorption\\
  15 & 16 & 9790 & 9980 & 10200 & 0.215 & Spline3/10 & 53 & 1.22 & low S/N\\
  \hline
  \hline
  \end{tabular}
\title{Calibration parameters}
\caption{Calibration parameters for UVB and VIS arm of \xs. The indices of the spectral orders as used in this paper are given in Col. 1, whereas the corresponding true spectral numbers of the dispersive element are given in Col. 2. Column 3 shows the minimum wavelength of each order, Col. 4 the wavelength at which the slit losses are estimated (close to the blaze wavelengths) and Col. 5 shows the maximum wavelength per order. The average dispersion is given in Col. 6, although the dispersion is a function of both slit and wavelength coordinate in each order. Column 7 lists the fitting functions and polynomial orders used in the sensitivity function computation. The number of underlying integration bandpasses is presented in Col. 8. The resulting RMS are shown in Col. 9; however, we note that values $<0.1$ and bandpass numbers $\lesssim 10$ are to be handled with care, as the fit is typically only weakly constrained in these cases. Finally, Col. 10 gives an overview of order-dependent features that might compromise the flux calibration accuracy.}
\label{tbl:sensfunc}
\end{table*}

\begin{figure*}[t]
\centering
\includegraphics[width=\linewidth]{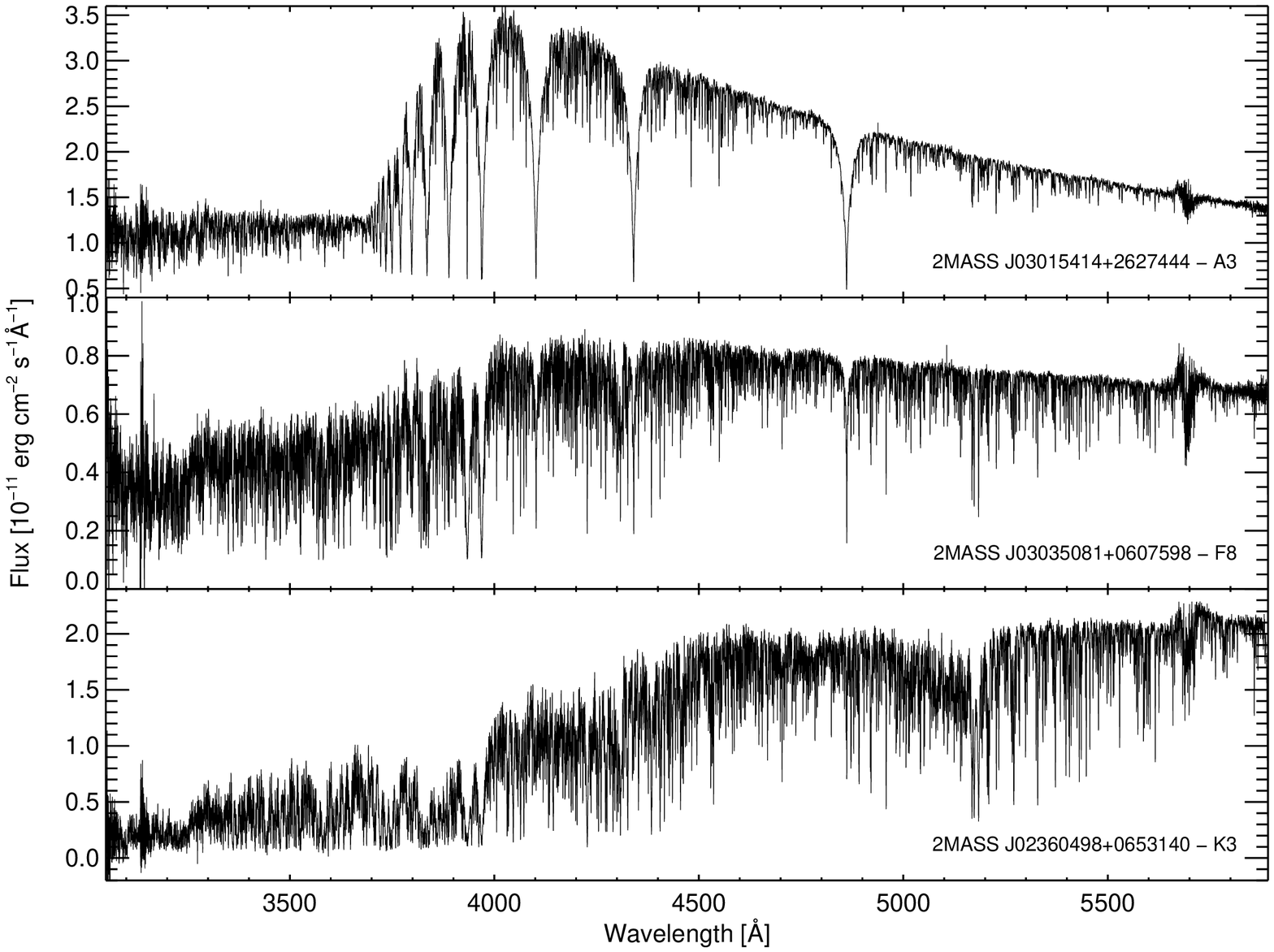}
\caption{Stellar spectra from ESO observing program 084.B-0869(A). All stars were observed on the night of 2009 November 26 and automatically extracted with the prescriptions presented in this work. Each target was observed in \emph{NOD} mode with a slit width of $0.5\arcsec$. The underlying sensitivity functions were derived with the spectrophotometric standard star GD 71. The three panels cover the spectral type range from a hot star with spectral type A3 ({\it top panel}), to an intermediate-type star F8 ({\it middle panel}), to a cool giant star with a spectral type K3 ({\it bottom panel}).} 
\label{fig:stellar_lib}
\end{figure*}

\subsubsection{Absolute flux calibration accuracy}
The middle panel of Figure \ref{fig:sens_func} shows the flux-calibrated output of our spectrophotometric reference observation (solid black line).~The S/N curve of the underlying data is above 50 per spectral bin, except for wavelengths $\lesssim\!3200\,\AA$. To highlight any deviations from the used SED model (overplotted in red), we show the ratio spectrum between data and model in the bottom panel of Figure~\ref{fig:sens_func}. For continuum regions, the systematic deviations between data and model are typically $\lesssim\!0.5\%$, with very localized increases to $\lesssim\!4\%$ at wavelengths where stellar features are present. Since atmospheric absorption bands occurring at wavelengths $\lesssim\!3500\,\AA$ and $\gtrsim\!6900 \,\AA$ have not been corrected for, the sensitivity functions are only weakly constrained in the affected wavelength regions. From our experience of calibrating a significant number of different spectrophotometric standard star observations with different fitting functions and polynomial orders, we estimate that a conservative upper limit for the uncertainty in the sensitivity functions in extended atmospheric windows is $10\%$. At wavelengths $\lesssim\!3050\,\AA$ the obtained S/N in a typical standard star observation is too low to derive a reliable sensitivity function and the flux calibration is limited by the low number of detected photons. 

According to \citet{bohlinmodel}, the systematic model $1\sigma$ uncertainties for GD71 range from $4\%$ at $1300\,\AA$ to $2\%$ between $5000\,\AA$ and $10000\,\AA$ for the stellar continuum and $5\%$ for the Balmer lines with respect to the adjacent continuum. Thus, the absolute flux calibration accuracy of wide-slit \xs observations will be mostly limited by model SED uncertainties as well as the photometric stability of the atmosphere. For narrow-slit spectra the uncertainties in the slit loss estimation (see Section \ref{sec:opt_extr}) will dominate the systematic error budget with a conservative upper limit of $5\%$ for $1.0\arcsec$ slit widths and $10\%$ for $0.5\arcsec$. Moreover, spectral ranges of low S/N ($\lesssim\!10$, mostly the blue parts of each order) might show a further degradation of the flux calibration accuracy by $2-3\%$, as illustrated in Figure \ref{fig:GD_71_slit_loss_correction}.~For all slit width setups we estimate that the absolute flux calibration is better than $\sim\!10$\% for the covered wavelength ranges of the UVB and VIS arm.

For demonstration purposes, we applied our reduction script to a sequence of stars with various spectral types.~We chose objects observed as part of the ESO observing program 084.B-0869(A) (PI: S. Trager)\footnote{The data were taken from the ESO data archive, accessible at http://archive.eso.org}, as they were taken with an instrumental setup that is similar to that of our dataset in the UVB arm.~As a consequence our reduction script can be executed automatically without any additional modifications.~The flux-calibrated output for three different stars (spectral types: A3, F8, K3), all observed on 2009 November 26, is presented in Figure \ref{fig:stellar_lib}. Each spectrum is a composition of two nodded frames, whose superposition has been optimally extracted and flux-calibrated against an observation of the spectrophotometric standard star GD 71 on the same night.~We additionally corrected for the effective airmass of each observation.~The S/N values are typically $\sim\!100$ per spectral bin, except for wavelengths $\lesssim 3500\,\AA$ and at the dip of the dichroic response at $\sim 5700\,\AA$. Because of the effects explained above, the latter region in particular has to be treated with great care when quantitative resultes are required.

\subsection{Fine-tuning of the wavelength calibration}
\label{sec:wavelength_update}
Since \xs is mounted at the Cassegrain focus of VLT, it suffers from flexure under its own weight of $\sim\!2.5$ metric tons \citep{dodorico}.~This effect is most pronounced for telescope pointings at large airmass and needs to be properly modeled and calibrated out if the full instrument potential is to be exploited.

\subsubsection{Wavelength scale update}
\begin{figure}[t]
\centering
\includegraphics[width=\linewidth]{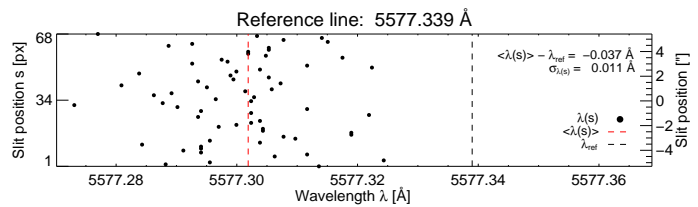}
\caption{Gaussian centroid wavelengths as a function of slit position for the prominent O\,{\sc i} line at $5577.339 \,\AA$. The median of all fitted wavelengths is shown with a vertical red dashed line, the corresponding air reference wavelength in dashed black. The difference between the measurement and reference is $-0.037\,\AA$ and the standard deviation of the fitted values is $0.011 \,\AA$. The cross-dispersed direction corresponds to the y-axis, the spectral dispersion direction to the x-axis.}
\label{fig:wavecal_single_line}
\end{figure}

Regular multi-pinhole ThAr arc exposures, which form the basis of the wavelength calibration, are only taken $\sim\!1\!-\!2$ times a week during daytime at zenith position, which means that extra frames have to be acquired to update the existing wavelength solution to the ambient parameters of the respective science exposures. For this reason, additional active flexure compensation (AFC) frames (short ThAr arc lamp exposures) are contemporaneously taken with each science observation, allowing for a line-shift measurement at the current telescope orientation with respect to the zenith position, yet there is only one valid reference line shift that can be reliably modeled in the UVB and VIS arm, as all other wavelength positions are subject to ADC corrections. This offset is then assumed to be a global shift of the wavelength solution of the respective arm. Although the instrument's spectral format is reported to be stable within $\sim\!1.5$ pixels \citep{moehlerphysmod, vernet2011} during a typical night, we implemented a quality control check to validate the global accuracy of the wavelength calibration for each science exposure. 

For this, we fit a set of $15$ non-blended telluric lines in the VIS arm with a Gaussian profile and compare the obtained centroid wavelengths to their respective reference (air) wavelengths\footnote{as given by http://www.pa.uky.edu/\~{}peter/atomic/}.~This procedure is performed individually for each position within in the slit, yielding for each telluric line both a median shift and standard deviation, reflecting the uncertainty of the fitting process. Figure~\ref{fig:wavecal_single_line} shows an exemplary case for the O\,{\sc i} transition at $5577.339 \,\AA$ for one of our observations of the globular cluster NGC\,7099, performed with the $0.5\arcsec$ wide slit. The difference between the median centroid wavelength of all fitted slit positions and the reference wavelength is $-0.037\,\AA$ and thus more than three times larger than the intrinsic uncertainty of the fitting process ($0.011\,\AA$). 

A global picture for the whole spectral range is presented in Figure \ref{fig:wavecal_all_lines}, where the residuals for the entire set of $15$ lines are plotted as a function of their respective wavelengths. Analyzing the weighted mean residual of all fitted wavelengths, we find it to be $\left<{\Delta \lambda}\right> = -0.034\,\AA$, with an average uncertainty of $\left<\sigma\right> = 0.022\,\AA$. This uncertainty can be interpreted as a conservative instrumental limit to the wavelength calibration, as -- on average -- individual lines cannot be modeled more accurately in this setup.~The best linear fit is overplotted in red and shows a minor slope of $-3.84\cdot 10^{-6}$. Despite its small amplitude, we want to emphasize that the bias on the residuals is observed in an optimal scenario in which the AFC measurements were conducted right before the corresonding science observation. As we have also encountered cases for which $\left|\left<\Delta \lambda\right>\right|\gtrsim 0.05\,\AA$, we suspect that there may be undiscovered systematic effects that are not taken into account by the \xs physical model and thus prevent a more accurate absolute wavelength calibration. 

\subsubsection{Correcting for residual flexure compensation drift}
To work around the issue of residual flexure compensation drift, we use the information provided by the axis intercept $\Delta\lambda_0$ and slope $\alpha$ of the fit to the residuals to de-bias the global wavelength solution. Since we want to avoid additional interpolation processes, we only change the wavelength grid by updating the appropriate header keywords. This is possible because of the linearity of the correction factor and the equidistance of the underlying wavelength grid. The corrected wavelength solution $\lambda_c$ becomes
\begin{equation}
 \lambda_c = (1-\alpha) \lambda_u - \Delta\lambda_0, 
 \label{eqn:wavecal_correction}
\end{equation}
where $\lambda_u$ are the nodes of the uncorrected wavelength grid. The new grid then has a new zero point \emph{CRVAL1} and stepping \emph{CDELT1}, and remains equidistant. 

\begin{figure}[t]
\centering
\includegraphics[width=\linewidth]{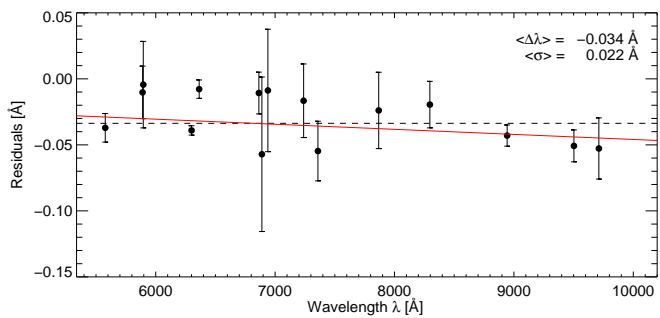}
\caption{Wavelength residuals as a function of wavelength for $15$ non-blended telluric lines. The individual data points and their uncertainties are derived as illustrated in Figure~\ref{fig:wavecal_single_line}. The global weighted mean residual is $-0.034\,\AA$ and is overplotted with a dashed black line. The error of the weighted mean is $0.022\,\AA$. The best-fit straight line is overplotted in red.}
\label{fig:wavecal_all_lines}
\end{figure}

\begin{figure}[b]
\centering
\includegraphics[width=\linewidth]{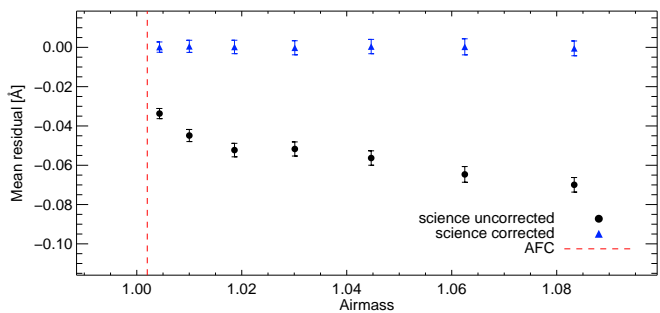}
\caption{Mean wavelength residuals as a function of airmass for six observations of NGC\,7099 and one additional sky frame. The black data points reflect the uncorrected weighted mean of all sky line residuals as explained in the text and shown in Figure \ref{fig:wavecal_all_lines}. Correcting the wavelength solution of each observation by means of Equation (\ref{eqn:wavecal_correction}) removes the bias (blue data points). The error bars illustrate the uncertainties of the weighted mean and represent the significance of the observed trend with airmass. The airmass at which the AFC measurement was performed is indicated with the dashed red line.}
\label{fig:wavecal_scans}
\end{figure}

A demonstration of this correction procedure is illustrated in Figure~\ref{fig:wavecal_scans}, which shows mean wavelength residual (of all 15 sky lines) vs.~airmass for the OB sequence of NGC\,7099, consisting of observations with various telescope positions. One additional sky observation was performed and the AFC frame was taken only once at the beginning of the sequence (dashed red line). Each frame was integrated for ten minutes with an overhead time of approximately two minutes in between. The residuals of the uncorrected wavelength solutions (black) show a clear drift with airmass in addition to the above mentioned bias of the wavelength solution of the initial science observation (the first data point in Figure \ref{fig:wavecal_scans} corresponds to the mean value inferred from Figure \ref{fig:wavecal_all_lines}). Correcting each science frame with its own correction factor based on Equation (\ref{eqn:wavecal_correction}) reliably removes the observed wavelength shift and corrects the wavelength solution (blue), resulting in spectra whose wavelength calibration is only limited by \xss internal uncertainty of $\sim\!0.02\,\AA$ as described above.  

Nevertheless, this correction method is currently only applicable to VIS arm data, as only in this spectral range can enough sky lines be robustly fitted. The UVB arm is physically decoupled from the VIS spectrograph and thus the computed correction factor is invalid. Additional calibration runs dedicated to this issue would need to be performed to investigate a possible correlation between the two arms. The NIR arm, however, should contain a sufficient number of usable sky lines so that the procedure described above should be applicable without great difficulties. The implementation and evaluation is intended for a future paper in this series.

\section{Summary and conclusions}
\label{sec:summary}
Reducing a large \xs dataset, such as our own observations of Local Group star clusters, comprising $\sim\!400$ raw and $\sim\!1500$ calibration frames, requires a careful analysis and precise understanding of both the data and the reduction processes. We implemented a sequence of advanced calibration steps wrapped around the existing ESO \xs pipeline (v1.5.0), which we use for the data extraction and rectification. The additional reduction procedures significantly improve the \xs data quality and integrity. In the following, we summarize our findings for the UVB and VIS arm of \xs.

For UVB arm data, we discovered a time-variable pick-up noise component on top of most of the analyzed frames; this component is perceived as an additional periodic pattern with an amplitude of about two counts.~Since the maxima are oriented along the \emph{NAXIS1} direction of the detector, we remove this feature for bias frames using a one-dimensional {\sc Fourier} filtering technique. The bias subtraction is then performed with a median stack of bias exposures that are free of pick-up noise.~For science data, this pick-up noise component can be modeled concurrently with the inter-order illumination background (mostly stray light) with one-dimensional polynomials, which are constrained at pixel positions between the curved echelle orders. Adjusting some of the CRH rejection parameters enables the ESO pipeline to flag and interpolate CRH affected pixel positions reliably without unwanted side effects. 

The wavelength calibration is derived with multiple ThAr arc frames and, with proper flexure compensation and an additional refinement based on telluric line measurements, a global accuracy of $0.02\,\AA$ can be achieved for narrow slit observations. Because of the implementation of the rectification process in the ESO pipeline, the wavelength-dependent dispersion relation of \xs has to be propagated to the rectified error frames. This requires a rescaling of the uncertainties of the rectified data to omit a constant S/N per output pixel -- independent of the specified output pixel size. This, however, does not account for the covariance introduced by the kernel convolution interpolation, which reduces the number of independent pixels by a factor of $\sim\!1.5$ and needs to be considered in any further analysis.

The bad pixel interpolation is performed on the rectified spectra. For this we propagate the bad pixel information through the rectification process and fit the rectified spectra with one-dimensional cubic splines along the cross-dispersed direction. The final correction is a weighted superposition of fit and data, where the weighting factors are based on the rectified bad pixel map. If a raw bad pixel is imaged one-to-one onto the rectified grid, the affected pixel value is replaced by the fit, whereas in cases in which, because of the kernel convolution, the rectified pixel position is only slightly affected by an adjacent bad neighbor, the interpolated data value dominates the superposition. 

To correct the rectified spectra for pixel-to-pixel variations we rectify a median stack of quartz lamp exposures and correct for the global structure of the measured count distribution. The utilized D2 lamp for the UVB orders $1\!-\!4$ shows a significant number of emission lines, which are removed by normalizing each wavelength bin by its median. As a matter of principle, flat fielding rectified spectra should ideally be omitted. However, our simulations show that the introduced uncertainties are $\lesssim\!0.5\%$ if compared to the accurate implementation of flat fielding the raw frame before rectification. For extended objects, we also recommend to account for systematic illumination inhomogeneities along the cross-dispersed direction. These are distinct on a $\pm2\%$ level for the narrowest slit widths and can be calibrated out with dedicated sky flat field frames. 

The sky subtraction for point-like objects is implemented with a spline-fit to two sky windows at either slit end of the rectified data and a subsequent interpolation to all slit positions. The break point stepping is adjusted to the smoothness of the sky signal and is set to two pixels for large gradients caused by sky lines and to eight pixels for sky continuum regions. The error propagation of this implementation is performed with a Monte Carlo simulation based on the rectified measurement uncertainties. For nodded data, the sky is naturally removed by shifting and co-adding the individually calibrated exposures. One dimensional data can be computed by selecting an optimal extraction aperture that optimizes the S/N of each wavelength bin. The resulting flux losses can be compensated with an analytical PSF model, based on a Moffat profile with an additional skewness factor to account for the observed instrumental distortions. To stabilize the fit in noisy scenarios the asymmetric component is penalized if the S/N drops below an adjustable threshold. With this technique, the S/N can be increased by a factor of $\sim\!1.5\!-\!2$, while conserving the flux on a $2\!-\!3\%$ level.~Moreover, a detailed understanding of the instrumental PSF allows for an accurate correction of flux losses at slit entry, with which the target flux can be reconstructed to an absolute uncertainty level of $5\!-\!10\%$ even for narrow slit widths of $\sim\!0.5\!-\!1.0\arcsec$. 

The sensitivity functions required for an absolute flux calibration can be derived from spectrophotometric standard star observations (typically obtained with $5.0\arcsec$ slit width) and their comparison to model SEDs. For continuum regions, the systematic errors in this procedure are typically dominated by model uncertainties ($2\!-\!4\%)$ and the stability of the atmosphere, but increase locally to $5\!-\!10\%$ in regions of spectral features or atmospheric absorption windows, where the smooth fit to the data is less constrained. The flux calibration accuracy in narrow slit setups will typically be limited by the reconstruction of slit losses and/or the low S/N of the data. Thus, with careful calibrations absolute flux calibration accuracy uncertainties of $\lesssim\!10\%$ can be achieved in virtually every scenario with a S/N $\gtrsim\!20$ per wavelength bin, which -- combined with an accuracy level of $0.02\,\AA$ for the wavelength calibration -- turns \xs into an extremely powerful instrument, whose unique capabilities will play a major role in the progress of astrophysical research in the current decade and beyond.

We aim to extend the calibrations presented in this work to \xss NIR arm in the future. Additionally, detailed telluric corrections for the atmospheric absorption bands at wavelengths $\gtrsim\!6900\,\AA$ (especially for NIR arm data) are intended. The corresponding data reduction scripts for the presented UVB and VIS arm data reduction procedures operate fully automatically, and are therefore suited to process large datasets in a consistent and rapid manner. The public release of the source code of the presented reduction scripts is planned for a future paper of this series.

\begin{acknowledgements}
FS acknowledges support by the {\it IMPRS} and {\it HGSFP}. FS thanks the Head of the {\it Office for Science of ESO Garching}, Eric Emsellem, for funding a fruitful visit to the X-shooter team at Garching.~Furthermore, FS is grateful for valuable visits at the {\it Institute of Astrophysics, Pontificia Universidad Cat\'olica de Chile}, where some of this work was conducted.~THP was supported by FONDECYT Regular Project Grant No.~1121005 and BASAL Center for Astrophysics and Associated Technologies (PFB-06).~THP is thankful for the hospitality and support during his visits at the {\it Astronomisches Rechen-Institut} of Heidelberg University, which were mainly funded by the visitor program of Collaborative Research Center ``The Milky Way System'' (SFB 881) of the German Research Foundation (DFG). AP and EKG acknowledge support by SFB 881 (subproject A5). We are grateful for inspiring and productive discussions with Robert Schmidt, Andrea Modigliani, Jo\"el Vernet, and Daniel Bramich, and we extend our gratitude to Christophe Martayan for providing special-purpose calibration datasets. We would also like to acknowledge the professional support of the ESO Paranal Observatory staff throughout the course of the observations, in particular Alvaro Alvarez, Mark Gieles, and Christophe Martayan, and also the telescope operators Angela Cortes, Nestor Jimenez, Patricia Guajardo, and Lorena Faundez. The achieved X-shooter data reduction quality and spectroscopic accuracy of the final science dataset would not have been possible without their invaluable and tireless support.~Last but not least, we thank our referee Philippe Prugniel for providing an insightful and constructive report that helped to improve the clarity of the manuscript and the methods presented in this work.
\end{acknowledgements}

\bibliographystyle{aa}
\bibliography{bibi}

\end{document}